\let\ifarxiv=\iftrue     
\pdfoutput=1
\documentclass[12pt,a4paper]{article}

\ifarxiv\ifnum\pdfoutput=1\else
\PassOptionsToPackage{hypertex}{hyperref}
\PassOptionsToPackage{draft}{graphicx}
\usepackage{showkeys}
\fi\fi

\usepackage{amsmath,amssymb}
\usepackage[bookmarks=true,hyperfigures=true]{hyperref}
\usepackage{graphicx}
\usepackage[nosort]{cite}
\usepackage[bulletsep]{collref}

\makeatletter
\newlength{\apb@width}
\newcommand{\autoparbox}[2][c]{\settowidth{\apb@width}{#2}\parbox[#1]{\apb@width}{#2}}
\newcommand{\includegraphicsbox}[2][]{\autoparbox{\includegraphics[#1]{#2}}}
\makeatother

\usepackage[a4paper,text={450pt,650pt},centering]{geometry}

\allowdisplaybreaks[3]

\numberwithin{equation}{section}

\usepackage[font=small,labelfont=bf,width=0.85\textwidth]{caption}

\makeatletter
\let\old@startsection=\@startsection
\renewcommand{\@startsection}[6]{\old@startsection{#1}{#2}{#3}{#4}{#5}{#6\mathversion{bold}}}
\makeatother

\let\oldPhi=\Phi
\let\oldPsi=\Psi
\let\oldGamma=\Gamma
\let\oldDelta=\Delta
\let\oldSigma=\Sigma
\let\oldLambda=\Lambda
\let\oldTheta=\Theta
\let\oldPi=\Pi
\let\oldXi=\Xi
\let\oldUpsilon=\Upsilon
\let\oldOmega=\Omega
\renewcommand{\Phi}{\mathnormal{\oldPhi}}
\renewcommand{\Psi}{\mathnormal{\oldPsi}}
\renewcommand{\Gamma}{\mathnormal{\oldGamma}}
\renewcommand{\Sigma}{\mathnormal{\oldSigma}}
\renewcommand{\Delta}{\mathnormal{\oldDelta}}
\renewcommand{\Theta}{\mathnormal{\oldTheta}}
\renewcommand{\Lambda}{\mathnormal{\oldLambda}}
\renewcommand{\Pi}{\mathnormal{\oldPi}}
\renewcommand{\Xi}{\mathnormal{\oldXi}}
\renewcommand{\Upsilon}{\mathnormal{\oldUpsilon}}
\renewcommand{\Omega}{\mathnormal{\oldOmega}}

\newcommand{\gen}[1]{\mathfrak{#1}}

\newcommand{\genY}[1]{\mathfrak{\widehat{#1}}}
\newcommand{\alg}[1]{\mathfrak{#1}}
\newcommand{\grp}[1]{\mathrm{#1}}
\newcommand{\superN}{\mathcal{N}}

\newcommand{\Tr}{\mathop{\mathrm{Tr}}}
\newcommand{\Dim}{\mathop{\mathrm{dim}}}
\newcommand{\tr}{\mathop{\mathrm{Tr}}}
\newcommand{\order}[1]{\mathcal{O}(#1)}

\newcommand{\ff}{f\kern-5pt f}

\newcommand{\dd}{d}
\newcommand{\eps}{\varepsilon}

\newcommand{\etaP}[1]{\eta_{#1}}
\newcommand{\partialP}[1]{\partial_{#1}}
\newcommand{\deltad}[1]{\delta^{#1}}

\ifx\genfrac\sdflkaj\else\fi
\newcommand{\sfrac}[2]{{\textstyle\frac{#1}{#2}}}
\newcommand{\half}{\sfrac{1}{2}}

\newcommand{\quarter}{\sfrac{1}{4}}

\newcommand{\indup}[1]{_{\mathrm{#1}}}

\newcommand{\supup}[1]{^{\mathrm{#1}}}
\newcommand{\matr}[2]{\left(\begin{array}{#1}#2\end{array}\right)}

\newcommand{\brk}[1]{(#1)}
\newcommand{\lrbrk}[1]{\left(#1\right)}
\newcommand{\bigbrk}[1]{\bigl(#1\bigr)}

\newcommand{\lrbrc}[1]{\left\{#1\right\}}

\newcommand{\comm}[2]{[#1,#2]}

\newcommand{\bigcomm}[2]{\bigl[#1,#2\bigr]}

\newcommand{\acomm}[2]{\{#1,#2\}}
\newcommand{\bigacomm}[2]{\bigl\{#1,#2\bigr\}}

\newcommand{\abs}[1]{|#1|}

\newcommand{\lreval}[1]{\left.#1\right|}

\newcommand{\tprod}[2]{\langle#1,#2\rangle}
\newcommand{\tprods}[2]{\langle#1#2\rangle}
\newcommand{\ctprod}[2]{[#1,#2]}
\newcommand{\ctprods}[2]{[#1#2]}
\newcommand{\bra}[1]{\langle #1|}
\newcommand{\ket}[1]{|#1\rangle}

\newcommand{\nn}{\nonumber}
\newcommand{\nln}{\nonumber\\}
\newcommand{\nl}[1][0pt]{\nonumber\\[#1]&\hspace{-4\arraycolsep}&\mathord{}}
\newcommand{\nlnum}{\\&\hspace{-4\arraycolsep}&\mathord{}}
\newcommand{\earel}[1]{\mathrel{}&\hspace{-2\arraycolsep}#1\hspace{-2\arraycolsep}&\mathrel{}}
\newcommand{\eq}{\earel{=}}
\newcommand{\beq}{\begin{equation}}
\newcommand{\eeq}{\end{equation}}

\def\[{\begin{equation}}
\def\]{\end{equation}}
\def\<{\begin{eqnarray}}
\def\>{\end{eqnarray}}

\makeatletter
\def\mr@ignsp#1 {\ifx\:#1\@empty\else #1\expandafter\mr@ignsp\fi}%
\newcommand{\multiref}[1]{\begingroup
\xdef\mr@no@sparg{\expandafter\mr@ignsp#1 \: }%
\def\mr@comma{}%
\@for\mr@refs:=\mr@no@sparg\do{\mr@comma\def\mr@comma{,}\ref{\mr@refs}}%
\endgroup}
\makeatother

\newcommand{\hypref}[2]{\ifx\href\asklfhas #2\else\href{#1}{#2}\fi}
\newcommand{\Secref}[1]{Section~\multiref{#1}}

\newcommand{\Appref}[1]{Appendix~\multiref{#1}}

\newcommand{\Figref}[1]{Figure~\multiref{#1}}

\renewcommand{\eqref}[1]{(\multiref{#1})}

\ifx\href\asklfhas\newcommand{\href}[2]{#2}\fi
\newcommand{\arxivlink}[1]{\href{http://arxiv.org/abs/#1}{arxiv:#1}}

\ifx\hypersetup\sadfkjashdfkxja\else
\hypersetup{pdftitle={Exacting N=4 Superconformal Symmetry}}
\hypersetup{pdfauthor={Till Bargheer, Niklas Beisert, Wellington Galleas, Florian Loebbert, Tristan McLoughlin}}
\hypersetup{pdfsubject={}}
\hypersetup{pdfkeywords={Scattering Amplitude, N=4 Supersymmetric Gauge Theory, Tree Level}}
\hypersetup{plainpages=false}
\hypersetup{pdfpagemode=UseNone}
\hypersetup{bookmarksnumbered=true}
\hypersetup{pdfstartview=FitH}
\hypersetup{colorlinks=false}
\hypersetup{citebordercolor={.5 1 .5}}
\hypersetup{urlbordercolor={.5 1 1}}
\hypersetup{linkbordercolor={1 .7 .7}}
\fi

\begin{document}

\thispagestyle{empty}
\ifarxiv\else
\begin{flushright}\footnotesize
\texttt{\arxivlink{0905.3738}}\\
\texttt{AEI-2009-048}%
\end{flushright}
\fi
\vspace{1cm}

\begin{center}%
{\Large\textbf{\mathversion{bold}%
Exacting $\mathcal{N}=4$ Superconformal Symmetry\ifarxiv$^\ast$\fi
}\par}
\vspace{1cm}%

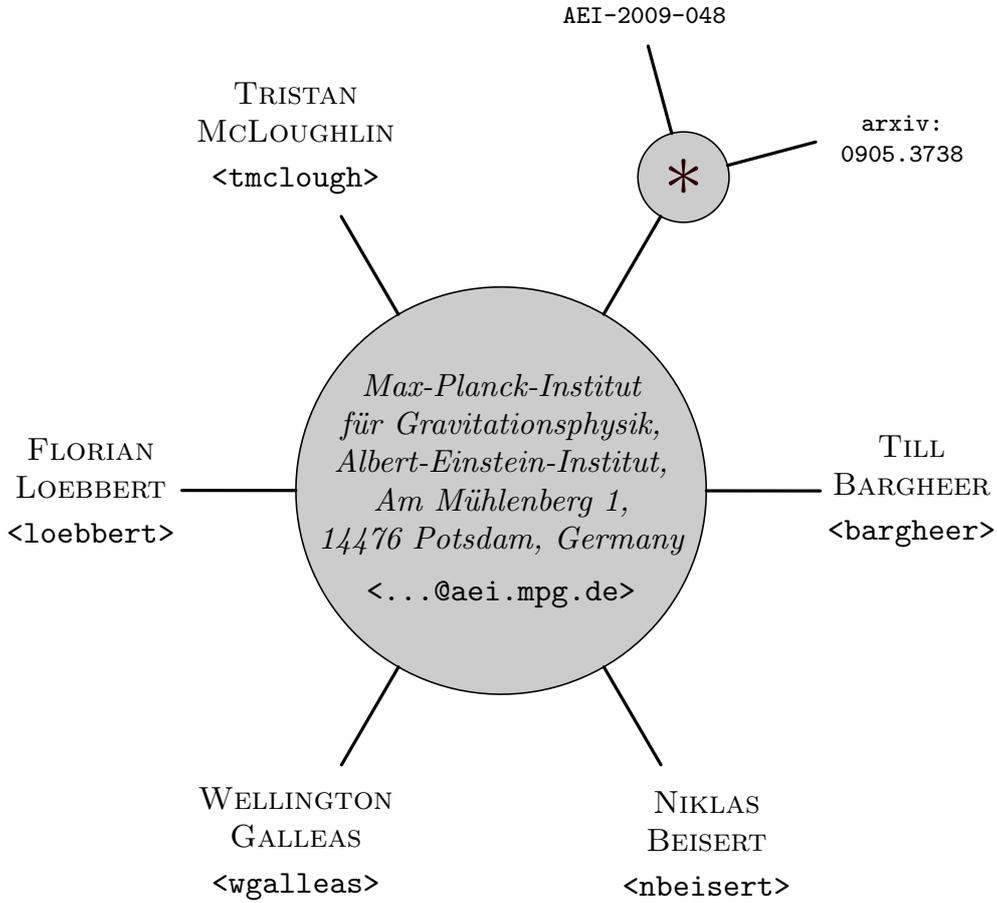
\ifarxiv\includegraphics[scale=1.2]{FigTitle.mps}\else
\textsc{Till Bargheer, Niklas Beisert, Wellington Galleas,\\Florian Loebbert and Tristan McLoughlin}\vspace{5mm}%

\textit{Max-Planck-Institut f\"ur Gravitationsphysik\\%
Albert-Einstein-Institut\\%
Am M\"uhlenberg 1, 14476 Potsdam, Germany}\vspace{3mm}%

\{\texttt{bargheer,nbeisert,wgalleas,loebbert,tmclough}\}\texttt{@aei.mpg.de}
\par\vspace{1cm}
\fi

\vspace{1cm}

\textbf{Abstract}\vspace{7mm}

\begin{minipage}{12.7cm}
Tree level scattering amplitudes in $\superN=4$ super Yang--Mills theory
are almost, but not exactly invariant under the free action 
of the $\superN=4$ superconformal algebra.
What causes the non-invariance is the 
holomorphic anomaly at poles where
external particles become collinear. 
In this paper we propose a deformation of the free superconformal representation 
by contributions which change the number of external legs. 
This modified classical representation not only makes tree amplitudes fully invariant,
but it also leads to additional constraints from symmetry alone
mediating between hitherto unrelated amplitudes. 
Moreover, in a constructive approach 
it appears to fully constrain all tree amplitudes 
when combined with dual superconformal alias Yangian symmetry.
\end{minipage}

\end{center}

\newpage

\setcounter{tocdepth}{2}
\hrule height 0.75pt
\tableofcontents
\vspace{0.8cm}
\hrule height 0.75pt
\vspace{1cm}

\setcounter{tocdepth}{2}

\section{Introduction and Overview}
\label{sec:Intro}

Maximally supersymmetric gauge theory in four spacetime dimensions
--- $\superN=4$ super Yang--Mills (SYM) ---
is an interacting quantum field theory with a host of useful features:
It has a unique massless action with only a few adjustable parameters.
Perturbative calculations typically show many 
cancellations such that, e.g., the model's classical conformal symmetry 
is preserved at the quantum level due to the absence of running couplings.
Furthermore, a lot of evidence has accumulated 
in favour of the AdS/CFT correspondence \cite{Maldacena:1998re,Gubser:1998bc,Witten:1998qj}
claiming that the model is exactly dual 
to a string theory on an AdS background. 

On top of these features, 
calculations in the planar alias the large-$N\indup{c}$ limit
for a $\grp{U}(N\indup{c})$ gauge group
have turned out to produce surprising final results in many cases.
Simplifications are certainly related to the absence of string interactions
in the dual string theory, yet it takes more to explain most of the observed mysteries.
Once fully understood and exploited, we hope that calculations 
at high perturbative orders and even at finite coupling become tractable.
For instance, the spectrum of anomalous dimensions of local operators
appears to be governed by a certain integrable model \cite{Minahan:2002ve,Beisert:2003tq,Bena:2003wd}
which makes calculations very efficient,
see e.g.\ the reviews \cite{Beisert:2004ry,Plefka:2005bk,Arutyunov:2009ga}.
Integrability is usually synonymous with the existence of 
an infinite dimensional algebra which enlarges the 
manifest symmetries of the model and which (almost) completely constrains 
the dynamics.
In this case superconformal symmetry apparently extends
to its loop algebra whose quantisation is a Yangian algebra 
\cite{Bena:2003wd,Dolan:2003uh}.

A different field of investigation in $\superN=4$ SYM which has 
advanced substantially in recent years is the study of on-shell scattering amplitudes. 
These are particularly important because of their relations 
to scattering amplitudes in QCD (for phenomenological purposes)
and in $\superN=8$ supergravity through the KLT relations
(for demonstrating finiteness of a particular theory of quantum gravity). 
In particular, the twistor space approach \cite{Witten:2003nn,Cachazo:2004kj} 
(see \cite{Mason:2009sa,ArkaniHamed:2009si,Hodges:2009hk} and references therein for further accounts)
following from the ideas of Penrose \cite{Penrose:1967wn}
has sparked many new investigations leading
to a much better understanding. 
Subsequently, recursion relations for all tree-level amplitudes have been set up 
\cite{Britto:2004ap,Britto:2005fq} 
and their on-shell superspace version 
\cite{Bianchi:2008pu,Brandhuber:2008pf,ArkaniHamed:2008gz,Elvang:2008na} 
solved explicitly \cite{Drummond:2008cr}.
Moreover, amplitudes at loop level can be computed 
efficiently and reliably through the methods of  generalised unitarity whose 
basic framework was introduced in 
\cite{Bern:1994zx,Bern:1994cg} and further developed in
\cite{Bern:1997sc,Britto:2004nc}; 
see \cite{Bern:2007dw} for a useful review.
Among others, these enabled the computation of the planar amplitudes 
with four legs up to four loops and beyond
\cite{Anastasiou:2003kj,Bern:2005iz,Bern:2006ew,Cachazo:2006az,Bern:2007ct}
as well as amplitudes with six or more legs at two loops 
\cite{Bern:2008ap,Vergu:2009zm}.

It is well known that scattering amplitudes for massless particles 
are problematic because asymptotic states cannot be defined properly:
a single massless particle can decay into an unbounded number 
of massless particles with collinear momenta.
This manifests itself in the appearance of infra-red divergences
at loop level when integrating over collinear momentum configurations. 
The divergences call for the introduction of some regulator, 
most commonly a minimal subtraction scheme in dimensional regularisation 
or reduction to $d=4-2\epsilon$ spacetime dimensions.
The resulting amplitudes will then have singularities
as $\epsilon\to 0$, typically two factors of $1/\epsilon$ per loop level.
The structure of IR divergences is understood reasonably well:
they combine into an exponent which can be factored out from the 
amplitude leaving a finite part behind 
\cite{Sen:1982bt, DelDuca:1989jt, Magnea:1990zb, Sterman:2002qn}. 
The form of the exponent is constrained by field theory 
and symmetry considerations. 
The same would be true for the finite remainder function,
however, some symmetries, such as special conformal transformations, 
may have been deformed or broken by the introduction of the regulator.

Some structural simplifications come about in the planar limit:
There the IR divergences are determined through a single function of the coupling, 
the so-called cusp anomalous dimension 
\cite{Korchemsky:1985xj,Korchemsky:1993hr,Korchemskaya:1994qp},
see also \cite{Bern:2005iz}.%
\footnote{The subleading collinear anomalous dimension 
is scheme dependent and not a good observable on its own.
For a recent discussion of these subleading singularities see \cite{Dixon:2008gr}.}
Interestingly, this very same cusp anomalous dimension
can also be computed from anomalous dimensions
of local operators which in turn are governed 
by the above mentioned integrable model,
see in particular \cite{Kotikov:2004er,Eden:2006rx,Beisert:2006ez}.
One might therefore wonder if there are further connections
between planar scattering amplitudes 
and the integrable structures for planar anomalous dimensions.

Indeed, the unitarity construction of higher-loop 
planar amplitudes shows some surprises: 
Many of the integrals that could in principle contribute
to the unitarity construction do not appear in practice. 
Only such integrals with certain conformal weights 
appear to have non-zero prefactors \cite{Drummond:2006rz,Bern:2006ew,Bern:2007ct}.
It is however not the standard conformal symmetry 
which leads to these restrictions, but rather
a conformal symmetry acting on momentum space.
Curiously, Wilson loops in this dual momentum space 
were seen to be equivalent to certain scattering amplitudes
\cite{Alday:2007hr,Alday:2007he,Drummond:2007aua,Brandhuber:2007yx,Drummond:2007cf,Drummond:2007au,Drummond:2007bm,Drummond:2008aq,Anastasiou:2009kna,McGreevy:2008zy,Komargodski:2008wa,Gorsky:2009nv},
see also the reviews \cite{Dixon:2008tu,Alday:2008yw,Henn:2009bd}.
Later the dual conformal symmetry was extended to 
superconformal symmetry and shown to apply to all tree level scattering amplitudes
\cite{Drummond:2008vq,Brandhuber:2008pf}.
In string theory the appearance of such dual superconformal 
symmetries can be explained by a supersymmetric
T-duality transformation which turns out to map the 
string model to itself \cite{Alday:2007hr,Berkovits:2008ic}.
The superconformal symmetries of the dual model become 
the dual superconformal symmetries of the original model.
Moreover, the two sets of superconformal symmetries 
form two inequivalent superconformal subalgebras 
of the loop algebra representing classical 
string integrability \cite{Beisert:2008iq,Berkovits:2008ic,Beisert:2009cs}.
Alternatively one can say that the
loop algebra alias integrability results as
the closure of the two sets of superconformal symmetries.
On the gauge theory side, the realisation of integrability 
alias Yangian symmetry for tree-level scattering amplitudes 
was derived in \cite{Drummond:2009fd} and shown to be self-consistent.

All of these developments together 
point towards integrability 
of planar scattering amplitudes in 
$\superN=4$ SYM, not only at tree level, 
but at all loops and even non-perturbatively. 
This suggests that one might be able to compute 
all planar scattering amplitudes
very efficiently and without the need for 
lengthy field theory or 
generalised unitarity calculations.
Could there be some differential or integral equation
determining the finite part of scattering amplitudes?

Before such an equation can be established, 
several problems have to be overcome: 
The regulator for the IR divergences breaks 
the special conformal symmetries.
E.g.\ in dimensional regularisation 
the dimensionality of spacetime is $d=4-2\epsilon$ 
while conformal symmetry requires exactly $d=4$.
Consequently conformal symmetry for scattering amplitudes 
is either broken beyond repair or it is at least obscured at loop level.
For dual conformal symmetry at loop level the second option 
seems to apply; its breakdown can be formulated as an anomaly
originating from UV divergences for the dual Wilson loops \cite{Drummond:2007au}. 
One may expect the same to be true for the original conformal symmetry.
The $\superN=4$ model is known to be exactly conformal at the quantum level. 
Conformal symmetry persists even in the presence of the UV divergences 
accompanying the anomalous dimensions of local operators. 
The main difficulty for scattering amplitudes
rests in the IR nature of the divergences
whose structure is less clear than for UV divergences.

The above discussion hides two important points at tree level
which appear to paint a pessimistic picture.
Firstly, conformal symmetry is subtle 
and even at tree level it does not strictly hold: 
Amplitudes were shown to be conformal when the external 
momenta are in a general position. Whenever two momenta
become collinear, however, conformal symmetry becomes anomalous.
A related anomaly is made  obvious by going to the  twistor space
representation of the amplitudes 
\cite{Mason:2009sa,ArkaniHamed:2009si}.%
\footnote{There are two subtleties here: 
a) the twistor space formulation requires the signature of
spacetime to be $(2,2)$ and not $(3,1)$; 
b) the twistor transformation itself is singular at collinearities.}
On a second thought this subtlety is not very surprising
because it is precisely the collinear momenta
which cause the IR divergences 
which in turn lead to the conformal anomaly. 
Only at tree level can collinearities be avoided through
a choice of external momenta
while at loop level internal momenta are integrated over 
and collinearities become inevitable.
Secondly, conformal and dual conformal 
symmetry together are not even sufficient 
to fix tree level amplitudes completely. 
The basis of tree amplitudes introduced in 
\cite{Drummond:2008cr} or similarly in the
twistor space picture is (almost, see above) 
invariant under both symmetries.
Consequently all linear combinations are invariant as well
and symmetry alone does not determine the correct linear combination
for the physical scattering amplitude.%
\footnote{We thank James Drummond, Johannes Henn and Emery Sokatchev 
for explanations.}
Only additional physical input, such as a correct set of singularities,
appears to fix the right coefficients, 
see also the very recent work \cite{Hodges:2009hk}
as well as \cite{Korchemsky:2009hm} which appeared
after an earlier version of the present work.

In this paper we propose a resolution to the problems 
of conformal symmetry at tree level discussed above: 
The naive action of infinitesimal conformal transformations 
on scattering amplitudes is not complete. 
It needs to be supplemented by correction terms
which cure the collinear anomaly at tree level. 
We also believe that similar corrections can
remove the anomalies at loop level and thus 
render scattering amplitudes exactly conformal, 
albeit using a deformed representation.
The proposed corrections act in similar fashion as the 
symmetry generators of the integrable spin chain
for anomalous dimensions. 
Most importantly, the corrections have the ability to change
the number of legs of scattering amplitudes.
Such generators cannot act on individual scattering amplitudes,
but rather they must act on the generating functional of all amplitudes. 

Altogether this paints a consistent picture in view of the problems 
introduced by massless asymptotic states:%
\footnote{We thank David Skinner for pointing this out and for discussions.}
The number of massless asymptotic particles is not a well-defined quantity.
Hence it is natural to consider the generating functional of scattering amplitudes 
(which can be viewed as the scattering operator)
rather than individual scattering amplitudes with a fixed number of legs.
The purpose of the correction terms is to take into account 
the overcounting of states in the Fock space where momenta become collinear.

The paper is organised as follows: 
We start in \Secref{sec:Conformal} by presenting how free superconformal symmetry acts on 
scattering amplitudes and compare it to the quantum action on local operators. 
We conclude that the action on amplitudes may require corrections
whose qualitative form is derived by analogy with local operators.
In \Secref{sec:exactMHV} we determine these corrections by 
demanding exact superconformal invariance of MHV amplitudes.
We then show the closure of the superconformal algebra 
modulo gauge transformations in \Secref{sec:Closure}.
Finally, in \Secref{sec:Trees} we show invariance of all
tree amplitudes under the deformed superconformal representation.
We summarise our results in \Secref{sec:Concl} and give an outlook.

\section{Representation of Superconformal Symmetry}
\label{sec:Conformal}

In this section we review and discuss the
representation of superconformal symmetry
on scattering amplitudes. 
By means of analogy to local operators
we propose how to qualitatively 
deform the free representation to 
an interacting one.

\subsection{Free Representation}

Scattering amplitudes in $\superN=4$ SYM are most conveniently 
expressed in the spinor helicity superspace \cite{Nair:1988bq}:
The light-like momentum $p$ of each external particle 
is first converted to a bi-spinor $p^{a\dot a}$
which can consequently be written as a product
$p^{a\dot a}=\lambda^a\bar\lambda^{\dot a}$. Here 
$\lambda^a$ and $\bar\lambda^{\dot a}$ are
mutually conjugate bosonic spinors of the Lorentz algebra
with $a,b,\ldots=1,2$ and $\dot a,\dot b,\ldots=1,2$.
The decomposition is unique up to a complex phase 
$\lambda^a\to e^{i\varphi}\lambda^a$
and $\bar\lambda^{\dot a}\to e^{-i\varphi} \bar\lambda^{\dot a}$.
Furthermore, it is advantageous to compute scattering amplitudes 
for the superfield \cite{Mandelstam:1982cb,Brink:1983pd}
\<
\Phi(\lambda,\bar\lambda,\eta)
\eq
G^+(\lambda,\bar\lambda)
+\eta^A \Gamma_A(\lambda,\bar\lambda) 
+\sfrac{1}{2}\eta^A\eta^B S_{AB}(\lambda,\bar\lambda) 
\nl
+\sfrac{1}{6} \varepsilon_{ABCD} \eta^A\eta^B \eta^C {\bar \Gamma}^D(\lambda,\bar\lambda) 
 +\sfrac{1}{24}\varepsilon_{ABCD}\eta^A\eta^B \eta^C\eta^D  G^-(\lambda,\bar\lambda),
\>
where $G^\pm$, $\Gamma$/$\bar\Gamma$, $S$ are the 
on-shell gluons, fermions and scalars with definite helicity.
By picking a suitable component in the expansion of 
fermionic spinors $\eta^A$, $A,B,\ldots=1,2,3,4$, of $\alg{su}(4)$ one can select 
the desired type of external particle for each leg.
The scattering amplitude for $n$ external particles
is thus a superspace function 
\[
A_n(\lambda_1,\bar\lambda_1,\eta_1,\ldots,\lambda_n,\bar\lambda_n,\eta_n).
\]

\begin{figure}\centering
\includegraphics{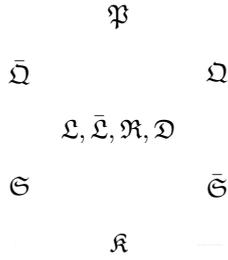}

\caption{Structure of the superconformal algebra $\alg{psu}(2,2|4)$.
The generators are plotted according to their 
scaling dimensions (vertical) and their helicities (horizontal).}
\label{fig:Superconformal}
\end{figure}

The superconformal algebra $\alg{psu}(2,2|4)$ can be represented in a simple fashion 
on such scattering amplitudes. We shall denote the 
superconformal generators through Gothic letters $\gen{J}_\alpha$.\ifarxiv\footnote{Resistance is futile.}\fi{}
More concretely, it is generated by Lorentz rotations $\gen{L},\gen{\bar L}$,
internal rotations $\gen{R}$, momentum generators $\gen{P}$, 
special conformal generators $\gen{K}$, the dilatation generator $\gen{D}$
as well as supercharges $\gen{Q}$, $\gen{\bar Q}$ and special conformal
supercharges $\gen{S}$, $\gen{\bar S}$, see \Figref{fig:Superconformal}.
Using the spinor helicity superspace coordinates
the representation of the superconformal algebra
can be written in a very compact fashion
\cite{Witten:2003nn} (cf.\ \Secref{sec:Closure})
\[\label{eq:StdRep}
\begin{array}[b]{@{}rclcrcl@{}}
\gen{L}^a{}_b\eq \lambda^a\partial_{b}-\half\delta^a_b \lambda^c\partial_c,
&&
\gen{\bar L}^{\dot a}{}_{\dot b}\eq 
\bar\lambda^{\dot a}\bar\partial_{\dot b}
-\half\delta^{\dot a}_{\dot b} \bar\lambda^{\dot c}\bar \partial_{\dot c},
\\[1ex]
\gen{D}\eq \half\partial_c\lambda^c+\half\bar\lambda^{\dot c}\bar \partial_{\dot c},
&&
\gen{R}^A{}_B\eq \eta^A\partial_B-\quarter \delta^A_B\eta^C\partial_C,
\\[1ex]
\gen{Q}^{aB}\eq \lambda^a\eta^B ,
&&
\gen{S}_{a B}\eq \partial_a\partial_B  ,
\\[1ex]
\bar{\gen{Q}}^{\dot a}_{B}\eq \bar\lambda^{\dot a} \partial_B ,
&&
\bar{\gen{S}}_{\dot a}^B\eq \eta^B \bar\partial_{\dot a} ,
\\[1ex]
\gen{P}^{a\dot b}\eq \lambda^a\bar\lambda^{\dot b} ,
&&
\gen{K}_{a\dot b}\eq \partial_a\bar\partial_{\dot b},
\end{array}
\]
where we abbreviate $\partial_a=\partial/\partial\lambda^a$, 
$\bar\partial_{\dot a}=\partial/\partial\bar\lambda_{\dot a}$ 
and $\partial_A=\partial/\partial\eta^A$.
Furthermore, let us introduce a central charge $\gen{C}$ and the 
helicity charge $\gen{B}$ which would extend the algebra to $\alg{u}(2,2|4)$. 
Their representation reads
\[
\gen{C}
=\partial_a\lambda^a-\bar\lambda^{\dot c}\bar \partial_{\dot c} - \eta^C\partial_C
=2+\lambda^a\partial_a-\bar\lambda^{\dot c}\bar \partial_{\dot c} - \eta^C\partial_C,
\qquad
\gen{B}= \eta^C\partial_C.
\]

In fact, this is only one half of the story:
The energy component in 
$p^{a\dot b}=\lambda^a\bar\lambda^{\dot b}$
is manifestly positive. 
However, reasonable scattering amplitudes require at least 
two particles with negative energy. 
For such particles we must set $p^{a\dot b}=-\lambda^a\bar\lambda^{\dot b}$.
The negative energy representation is the same as the above \eqref{eq:StdRep},
where the sign of all instances of $\bar\lambda$ is flipped.
In most places this replacement is sufficient and can be done mechanically. 
We shall thus treat all particles as though their energy is positive
and point out whenever negative energy particles make an essential difference (cf.\ \Secref{sec:NegEn}).

The representation on tree-level scattering amplitudes in $\superN=4$ SYM 
takes the standard tensor product form 
\[\label{eq:RepLO}
\gen{J}_\alpha=\sum_{k=1}^n \gen{J}_{k,\alpha}.
\]
Here $\gen{J}_{k,\alpha}$ is the representation of the 
conformal symmetry generator $\gen{J}_\alpha$ 
on the $k$-th leg $(\lambda_k,\bar\lambda_k,\eta_k)$ 
of $A_n$ as specified in \eqref{eq:StdRep}.
Invariance of $A_n$ is the statement
\[
\gen{J}_\alpha A_n=0.
\]

In \cite{Drummond:2009fd} a Yangian representation 
on tree-level scattering amplitudes in $\superN=4$ SYM was proposed.
The action of the level-one Yangian generators $\genY{J}_\alpha$ 
follows the standard Yangian coproduct rule
for evaluation representations with homogeneous evaluation parameters
\[\label{eq:RepYang}
\genY{J}_\alpha=\half f_{\alpha}^{\beta\gamma}\sum_{1\leq k<\ell\leq n} \gen{J}_{k,\beta}\gen{J}_{\ell,\gamma}.
\]
This representation was shown to be compatible with cyclicity
provided that the amplitude is invariant under superconformal symmetry.
Making use of dual superconformal covariance \cite{Drummond:2008vq,Brandhuber:2008pf}
and the Serre relations
one can further deduce that the tree-level amplitudes
are invariant under the complete Yangian algebra,
$\genY{J}_\alpha A_n=0$ \cite{Drummond:2009fd}.

\subsection{Higher-Loop Representation on Local Operators}

\begin{figure}
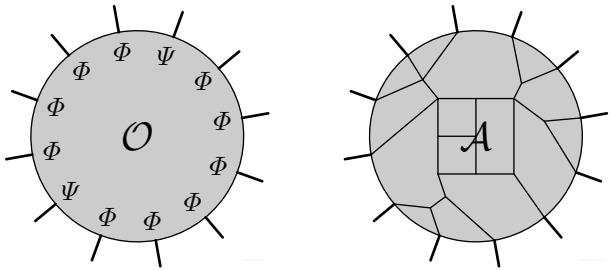
\centering
\includegraphics{FigCompareSpinChain.mps}
\qquad
\includegraphics{FigCompareAmplitude.mps}

\caption{Comparison of single-trace local operators and colour-ordered scattering amplitudes}
\label{fig:LocalScat}
\end{figure}

This representation is the direct analog of the 
leading-order representation on local operators when $\gen{J}_{k,\alpha}$
is the representation on the $k$-th site of the spin chain.
The main difference is that $\gen{J}_{k,\alpha}$ is a differential operator
for scattering amplitudes while it is a spin operator for local operators.%
\footnote{Without going into details, 
the oscillator representation introduced in \cite{Gunaydin:1984fk}
and applied in \cite{Beisert:2003jj} is practically
equivalent to the above representation.}
In fact, the structures of single-trace local operators and
colour-ordered scattering amplitudes are very much alike
as illustrated in \Figref{fig:LocalScat}.

It is well-known that the representation of the superconformal 
algebra on local operators is deformed at loop level. 
This is required to incorporate the effects of anomalous dimensions;
after all the dilatation generator measures conformal dimensions.
Alternatively one can say that the deformation is due to 
regularisation of UV divergences. 
While the tree-level generators $\gen{J}_{k,\alpha}$
act on a single site of the local operator 
and map it back to itself, 
the structure of the loop corrections is qualitatively different: 
They can act on several sites at the same time 
and map them back to themselves. 
Moreover, they are \emph{dynamic} in the sense that 
they can change the number of sites, 
e.g.\ map a single site to two sites or vice versa \cite{Beisert:2003ys}.
This implies that local operators with well-defined scaling dimension 
do not have a well-defined number of component fields, 
but they are rather linear combinations of spin chains with different lengths.
Note that some of these length-changing effects 
are known as ``non-linear'' or interacting realisations of the symmetry. 
For example, it is well-known that a supercharge $\gen{Q}$ acting on 
a fermion can produce the commutator of two scalars \cite{Beisert:2003ys}
even in the classical theory.

\begin{figure}
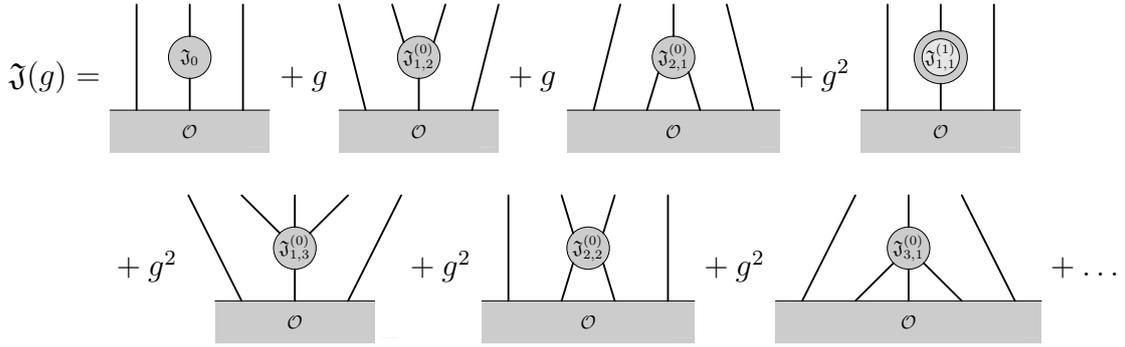
\centering
$\begin{array}{rcl}
\gen{J}(g)\eq
    \includegraphicsbox[scale=0.7]{FigChainJ0.mps}
+g\ \includegraphicsbox[scale=0.7]{FigChainJ1a.mps}
+g\ \includegraphicsbox[scale=0.7]{FigChainJ1b.mps}
+g^2\ \includegraphicsbox[scale=0.7]{FigChainJ2c.mps}
\\[7ex]\earel{}\mathord{}
+g^2\ \includegraphicsbox[scale=0.7]{FigChainJ2a.mps}
+g^2\ \includegraphicsbox[scale=0.7]{FigChainJ2b.mps}
+g^2\ \includegraphicsbox[scale=0.7]{FigChainJ2d.mps}
+\ldots
\end{array}$
\caption{Expansion of quantum symmetry generators for local operators.}
\label{fig:Dynamic}
\end{figure}

The generic structure of the perturbative representation $\gen{J}_\alpha(g)$ for some generator 
$\gen{J}_\alpha$ 
around the free representation $(\gen{J}_0)_\alpha=(\gen{J}^{(0)}_{1,1})_\alpha$
reads
\[\label{eq:Dynamic}
\gen{J}_\alpha(g)=\sum_{m,n=1}^\infty\sum_{\ell=0}^\infty 
g^{2\ell+m+n-2}(\gen{J}^{(\ell)}_{m,n})_{\alpha}.
\]
This structure follows from the structure of planar Feynman graphs \cite{Beisert:2003tq,Beisert:2003ys},
and it is depicted in \Figref{fig:Dynamic}.
An $\ell$-loop contribution $\gen{J}^{(\ell)}_{m,n}$ 
which acts on $m$ adjacent sites of the chain 
and which replaces them by $n$ adjacent sites
is of order $g^{2\ell+m+n-2}$.
This is because an elementary interaction of $\order{g}$ 
connects three sites; adjacency is due to the planar limit.

\subsection{Higher-Loop Representation on Scattering Amplitudes}

Now one could imagine that similar deformations 
apply to the representation of conformal symmetry on scattering amplitudes.
Clearly the origin of the corrections is different: 
for local operators it is due to UV divergences 
whereas for scattering amplitudes it is due to IR divergences. 
This means that perhaps the representations are not exactly equivalent. 
Nevertheless one would expect the structural constraints to be the same
because they merely originate from the structure of Feynman graphs
(and the planar limit).

The action of deformations which involve several legs, but 
preserve their number should be self-evident. 
But what does it mean to change the number of legs? 
In particular, how can this be possible at all if each leg
has a well-defined particle momentum?
How can invariance of a scattering amplitude be interpreted?
First of all, if the number of legs changes by the action of 
symmetry generators, then a single $n$-leg amplitude cannot be invariant by itself;
it only makes sense to talk about invariance 
of all amplitudes at the same time.

\begin{figure}
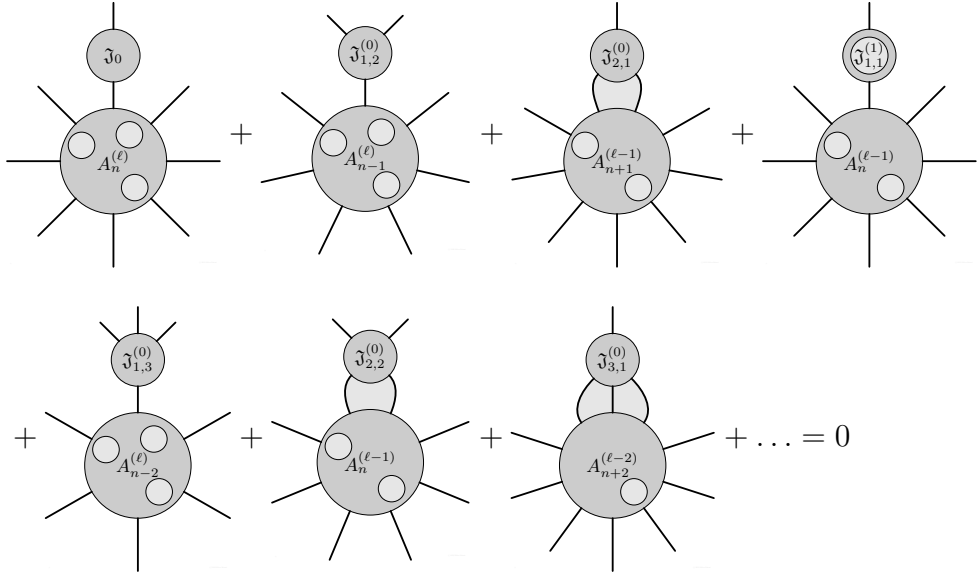
\centering
$\begin{array}{l}
\includegraphicsbox[scale=0.7]{FigAmpJ0.mps}
+\includegraphicsbox[scale=0.7]{FigAmpJ1a.mps}
+\includegraphicsbox[scale=0.7]{FigAmpJ1b.mps}
+\includegraphicsbox[scale=0.7]{FigAmpJ2c.mps}
\\[11ex]\mathord{}
+\includegraphicsbox[scale=0.7]{FigAmpJ2a.mps}
+\includegraphicsbox[scale=0.7]{FigAmpJ2b.mps}
+\includegraphicsbox[scale=0.7]{FigAmpJ2d.mps}
+\ldots
=0
\end{array}$
\caption{Expansion of quantum invariance of scattering amplitudes.
Loops (light grey) can appear inside the amplitudes,
inside the symmetry generator or in the connection of the two.}
\label{fig:DynamicInvariance}
\end{figure}

Before we introduce a proper framework for the treatment of 
length-changes, let us discuss their effects qualitatively.
Suppose a generator consists of the terms depicted in \Figref{fig:Dynamic}
\[\label{eq:DynamicGeneratorEx}
\gen{J}(g)=
\gen{J}_0
+g\gen{J}^{(0)}_{1,2}
+g\gen{J}^{(0)}_{2,1}
+g^2\gen{J}^{(1)}_{1,1}
+g^2\gen{J}^{(0)}_{1,3}
+g^2\gen{J}^{(0)}_{2,2}
+g^2\gen{J}^{(0)}_{3,1}
+\ldots
\]
The first term is the free generator $\gen{J}_0=\gen{J}^{(0)}_{1,1}$. 
The contributions
$\gen{J}^{(0)}_{1,2},\gen{J}^{(0)}_{1,3}$ increase the number of legs by one or two, respectively,
while $\gen{J}^{(0)}_{2,1},\gen{J}^{(0)}_{3,1}$ decrease it.
The symbol $\gen{J}^{(1)}_{1,1}$ represents the loop correction to the free generator
and $\gen{J}^{(0)}_{2,2}$ maps two legs to two legs. 
Suppose further that the set of amplitudes can be written 
as the linear combination 
\[\label{eq:AmplitudeEx}
\mathcal{A}(g)
=\sum_{n=4}^\infty g^{n-2} A_n(g)
=\sum_{n=4}^\infty\sum_{\ell=0}^\infty g^{n-2+2\ell} A^{(\ell)}_n
.
\]
Note that we have included a factor of $g$ for each three-vertex 
in the underlying Feynman graph;
this counting is compatible with the counting of $g$ in the expansion 
of $\gen{J}(g)$.
Demanding invariance of all amplitudes, $\gen{J}(g)\mathcal{A}(g)=0$,
and separating the terms according to their number of external legs
as well as the power of $g$ leads to the following invariance equation
\[\label{eq:DynamicInvarianceEx}
 \gen{J}^{(0)}_{1,1} A^{(\ell)}_n
+\gen{J}^{(0)}_{1,2} A^{(\ell)}_{n-1}
+\gen{J}^{(0)}_{2,1} A^{(\ell-1)}_{n+1}
+\gen{J}^{(1)}_{1,1} A^{(\ell-1)}_n
+\gen{J}^{(0)}_{1,3} A^{(\ell)}_{n-2}
+\gen{J}^{(0)}_{2,2} A^{(\ell-1)}_n
+\gen{J}^{(0)}_{3,1} A^{(\ell-2)}_{n+2}
+\ldots
=0.
\]
In particular the illustration of this 
equation in \Figref{fig:DynamicInvariance}
shows that the loop counting 
includes loops within the \emph{amplitude},
loops within the \emph{symmetry generator}
as well as loops formed by \emph{connecting} the two.

Note that it makes sense to rescale $n$-leg amplitudes
$A_n$ by a factor of $g^{2-n}$
such that all tree amplitudes are at $\order{g^0}$
and such that $g^2$ exclusively counts the number
of loops in Feynman graphs.
Thus we would use instead of \eqref{eq:AmplitudeEx}
and \eqref{eq:Dynamic}
\[
\mathcal{A}(g)=\sum_{n=4}^\infty A_n(g)
=\sum_{n=4}^\infty\sum_{\ell=0}^\infty g^{2\ell} A^{(\ell)}_n
,
\qquad
\gen{J}_\alpha(g)=\sum_{m,n=1}^\infty\sum_{\ell=0}^\infty 
g^{2(\ell+m-1)}(\gen{J}^{(\ell)}_{m,n})_{\alpha}.
\]
This is the normalisation that we shall use in the present work.
Note that this leads to the same invariance equation \eqref{eq:DynamicInvarianceEx}.

The crucial observation one can make in \eqref{eq:DynamicInvarianceEx} 
is that generators which act on a single leg and replace it by several legs,
such as $\gen{J}^{(0)}_{1,n}$, contribute to the same order as the free generator $\gen{J}_{0}$.
The conclusion would be that tree amplitudes in $\mathcal{A}(0)$
are not invariant under $\gen{J}_0$, 
but rather under the combination $\gen{J}_\alpha(0)$:
\[\label{eq:TreeInvariance}
\gen{J}_\alpha(0)\mathcal{A}(0)=0
\qquad\mbox{with}\qquad
\mathcal{A}(0)=\sum_{n=4}^\infty A^{(0)}_n
,\qquad
\gen{J}_\alpha(0)=\sum_{n=1}^\infty
(\gen{J}^{(0)}_{1,n})_{\alpha}.
\]
On the one hand this type of invariance is reasonable 
because terms like $\gen{J}^{(0)}_{1,2}+\ldots$ are precisely the ``non-linear''
contributions to symmetries in the interacting \emph{classical} theory.
Naturally these would have to be the proper symmetries for 
tree level amplitudes and not their free truncations $\gen{J}_0$.
On the other hand, tree-level amplitudes at first sight
do seem to be invariant under the free generators. 
Therefore the interacting correction terms $\gen{J}^{(0)}_{1,2}+\ldots$ would
either have to be trivial or they would have to annihilate the amplitudes 
on their own and independently of $\gen{J}_0$.
Both alternatives are somewhat unsatisfactory and indeed
there is a third: Tree-level amplitudes are 
\emph{not invariant} under the free action $\gen{J}_0$ of the symmetry. 
This violation of conformal symmetry 
is subtle and therefore is not immediately seen.
For generic external momenta the amplitudes are indeed invariant 
under naive conformal symmetry.
However, when the amplitudes are treated as distributions,
the action of $\gen{J}_0$ leaves certain contact terms 
when two adjacent momenta become collinear. 
Collinearity is essential because breaking up one massless particle
into two by means of $\gen{J}^{(0)}_{1,2}+\ldots$ can only produce collinear momenta
due to momentum conservation. 
In conclusion, it is conceivable that conformal symmetry
has a representation under which the tree-level amplitudes
are exactly invariant in a distributional sense. 
In particular, the length-changing effects would be crucial
for this representation. 
It would also be the proper starting point for extending 
the symmetries to the loop level.

\subsection{Amplitude Generating Functional}
\label{sec:AmpGen}

Before we consider concretely the length-changing contributions 
we shall first introduce a framework to deal with such terms.

\begin{figure}\centering
$\displaystyle
\mathcal{A}[J]=
    \frac{1}{4}\includegraphicsbox[scale=0.7]{FigAmpGen4.mps}
+   \frac{1}{5}\includegraphicsbox[scale=0.7]{FigAmpGen5.mps}
+   \frac{1}{6}\includegraphicsbox[scale=0.7]{FigAmpGen6.mps}
+   \frac{1}{7}\includegraphicsbox[scale=0.7]{FigAmpGen7.mps}
+\ldots\,.
$
\caption{The generating functional of colour-ordered scattering amplitudes.
The prefactors $1/n$ are the appropriate symmetry factors 
for cyclicity of the trace.}
\label{fig:AmpGen}
\end{figure}

On a technical level we can combine all scattering amplitudes
into a single generating functional.
Let $J(\lambda,\bar\lambda,\eta)$ be a source field 
corresponding to the superspace field $\Phi(\lambda,\bar\lambda,\eta)$.
For clarity of notation we shall combine the bosonic and fermionic superspace coordinates 
into a single symbol $\Lambda=(\lambda^{a},\bar\lambda^{\dot a},\eta^A)$.
The superspace measure is given through
$d^{4|4}\Lambda:=d^4\lambda\,d^4\eta$, 
see \Appref{app:Conventions}.
The generating functional $\mathcal{A}$ 
of colour-ordered amplitudes $A_n$
then reads simply, cf.\ \Figref{fig:AmpGen} 
(see also\ \cite{ArkaniHamed:2009si})
\[\label{eq:AmpGen}
\mathcal{A}[J]
=
\sum_{n=4}^\infty
\int 
d^{4|4}\Lambda_1\, \ldots d^{4|4}\Lambda_n\, 
\frac{1}{n}\Tr \bigbrk{J(\Lambda_1)\ldots J(\Lambda_n)}\,
A_n(\Lambda_1,\ldots,\Lambda_n).
\]
Conversely, the $n$-particle amplitude can be extracted as the variation
\[
A_n(\Lambda_1,\ldots,\Lambda_n)=
\lreval{
\frac{1}{N\indup{c}^n}
\Tr 
\lrbrk{
\frac{\delta}{\delta J(\Lambda_n)}\,
\ldots 
\frac{\delta}{\delta J(\Lambda_1)}}
\mathcal{A}[J]}_{J=0}.
\]
Note that the traces incorporate the colour structure of
colour-ordered amplitudes and $1/n$ is the proper symmetry factor.

For representations of $\alg{psu}(2,2|4)$
the central charge of $\alg{su}(2,2|4)$ must act trivially. 
This implies that the fields $\Phi(\lambda,\bar\lambda,\eta)$
are homogeneous functions under a simultaneous phase shift
of the arguments
\[\label{eq:PhiHom}
\Phi(e^{i\varphi}\Lambda)=e^{-2i\varphi}\Phi(\Lambda),
\qquad
e^{i\varphi}\Lambda:=(e^{i\varphi}\lambda,e^{-i\varphi}\bar\lambda,e^{-i\varphi}\eta).
\]
Consequently, the same must be true for each leg of the amplitude,
$A_n(\ldots,e^{i\varphi}\Lambda_n,\ldots)=
e^{-2i\varphi}A_n(\ldots,\Lambda_n,\ldots)$.
The Jacobian of the measure also leads to a weight
$d^{4|4}(e^{i\varphi}\Lambda)=e^{4i\varphi} d^{4|4}\Lambda$.
We will not impose a homogeneity condition for the source fields $J(\Lambda)$ 
so that the variations $\delta/\delta J$ can be performed straight-forwardly.
It is nevertheless clear that the generating functional \eqref{eq:AmpGen}
projects to the part of $J$ with definite scaling
\[\label{eq:Jhat}
\hat{J}(\Lambda):=\frac{1}{2\pi}\int_{0}^{2\pi} d\varphi\, e^{2i\varphi} J(e^{i\varphi}\Lambda).
\]
Therefore each factor of $J$ in \eqref{eq:AmpGen} can safely be replaced by $\hat J$;
the integral over $d^{4|4}\Lambda$ contains a similar integral over $d\varphi$.
Note that the projection $\hat J$ turns out to have the same homogeneity 
as $\Phi$ in \eqref{eq:PhiHom}.

In the framework of the generating functional 
the superconformal generators \eqref{eq:StdRep}
take the form of variations,
cf.\ \cite{Beisert:2002ff,Beisert:2003tq} for a similar representation.
For convenience we shall abbreviate variation by an accent $\check J$ on
the field $J$
\[
\check{J}(\Lambda):=\frac{\delta}{\delta J(\Lambda)}\,.
\]
Here we list only a few of the relevant generators
\[\label{eq:genfunc}
\begin{array}[b]{@{}rclrcl@{}}
\displaystyle (\gen{Q}_0)^{aB}\eq\displaystyle
\int d^{4|4}\Lambda\,
\Tr \lambda^a\eta^B J(\Lambda)\, \check{J}(\Lambda) \,,
&
\displaystyle (\gen{S}_0)_{a B}\eq\displaystyle
\int d^{4|4}\Lambda\,
\Tr \partial_a\partial_B J(\Lambda)\, \check{J}(\Lambda) \,,
\\[2ex]
\displaystyle (\bar{\gen{Q}}_0)^{\dot a}_{B}\eq\displaystyle
-\int d^{4|4}\Lambda\, 
\Tr \bar\lambda^{\dot a} \partial_B J(\Lambda)\, \check{J}(\Lambda) \,,
&
\displaystyle (\bar{\gen{S}}_0)_{\dot a}^B\eq\displaystyle
-\int d^{4|4}\Lambda\, 
\Tr \eta^B \bar\partial_{\dot a} J(\Lambda)\, \check{J}(\Lambda) \,,
\\[2ex]
\displaystyle (\gen{P}_0)^{a\dot b}\eq\displaystyle
\int d^{4|4}\Lambda\, 
\Tr \lambda^a\bar\lambda^{\dot b} J(\Lambda)\, \check{J}(\Lambda) \,,
&
\displaystyle (\gen{K}_0)_{a\dot b}\eq\displaystyle
\int d^{4|4}\Lambda\,
\Tr \partial_a\bar\partial_{\dot b}J(\Lambda)\, \check{J}(\Lambda)  \,.
\end{array}
\]
After performing the variations on $\mathcal{A}$
the derivatives should be integrated by parts to 
make them act on $A_n$ as for \eqref{eq:StdRep}.
A classical contribution to add one leg
takes the qualitative form $\gen{J}^{(0)}_{(1,2)}\sim \int \Tr JJ\check{J}$.
In practice, it acts by taking away one source term and replacing it by two.
The precise form of such contributions will be worked out in the 
following section.

Let us note that the above expressions for the superconformal generators 
remain valid even for amplitudes without colour ordering and 
away from the planar limit or for generic gauge groups.
Also the classical length-changing contributions are expected 
to remain valid at finite $N\indup{c}$.
Conversely, a representation of the Yangian cannot be formulated using 
the generating functional because one needs a framework which can make explicit
reference to specific legs, e.g.\ legs $k$ and $\ell$ as in \eqref{eq:RepYang}.

\section{Superconformal Invariance of MHV Amplitudes}
\label{sec:exactMHV}

In this section we wish to use the known form of the 
tree level MHV amplitudes to determine the necessary
deformations $\gen{J}^{(0)}_{1,n}$ of the classical conformal symmetry
generators (cf.\ \eqref{eq:TreeInvariance}). 

\subsection{MHV Amplitudes}
\label{sec:MHV}

Scattering amplitudes can be classified through their helicity. 
It is measured by the generator $\gen{B}$
counting the number of $\eta$'s
\[
A_n=\sum_{k=2}^{n-2}A_{n,k},\qquad
 \gen{B} A_{n,k}=4k A_{n,k}.
\]
The number of $\eta$'s ranges between $8$ for MHV
amplitudes and $4n-8$ for $\overline{\mbox{MHV}}$ amplitudes
\[
A\supup{MHV}_n=A_{n,2},\qquad
A\supup{\overline{MHV}}_n=A_{n,n-2}.
\]

The tree level MHV amplitudes of $\mathcal{N}=4$ SYM 
have a simple form when written in terms of Lorentz invariant
products of  spinors \cite{Parke:1986gb,Berends:1987me}
and  particularly so in the manifestly
supersymmetric formulation \cite{Nair:1988bq} using the
appropriate on-shell superspace. 
They take the form%
\footnote{We neglect an overall factor of $i (2\pi)^4$ in our definition 
of the amplitudes and similarly they are normalised so that
there is no prefactor of the coupling.}
\[\label{eq:MHV}
A\supup{MHV}_n
=
\frac{\deltad{4}(P)\,\deltad{8}(Q)}{\tprods{1}{2}\tprods{2}{3}\ldots \tprods{n}{1}}\,,
\qquad
P^{a\dot b}=\sum_{k=1}^n \lambda^a_{k}\bar\lambda^{\dot b}_{k}\,,
\quad
Q^{aB}=\sum_{k=1}^n \lambda^a_{k}\eta^B_{k}\,,
\]
where the brackets are defined in \Appref{app:Conventions}.

In the physically relevant case the spacetime signature is $(3,1)$.
It implies that the above expression for the amplitude cannot be entirely meaningful
because it assumes that all particles have strictly positive energies 
while energy conservation requires the sum of all energies to vanish.
For non-trivial amplitudes 
at least two particles should have negative energies. 
A particle $k$ with negative energy is achieved by flipping the sign of $\bar\lambda_k$. 
For the time being we shall ignore the implications of overall momentum conservation 
and assume all energies to be positive. The minute modifications due to 
negative energy particles will be discussed in \Secref{sec:NegEn}.

We now act with the free superconformal generator 
$(\gen{\bar S}_0)^B_{\dot a}=\sum_{k=1}^n\eta^B_k\bar\partial_{k,\dot a}$
as defined in \eqref{eq:StdRep} on the above amplitude.
Except for the delta function, the amplitude is holomorphic in the $\lambda_k$.
Thus, at first sight, the generator seems to act only on the delta function
\[
(\gen{\bar S}_0)^B_{\dot a}\deltad{4}(P)
=
\sum_{k=1}^n\eta^B_k\,\frac{\partial}{\partial \bar\lambda^{\dot a}_{k}}\,
\deltad{4}(P)
=
\sum_{k=1}^n\eta^B_k\lambda_k^c\,
\frac{\partial \deltad{4}(P)}{\partial P^{c\dot a}}
=
Q^{cB}\,
\frac{\partial \deltad{4}(P)}{\partial P^{c\dot a}}\,.
\]
The fermionic delta function $\deltad{8}(Q)$ ensures that the action vanishes 
$\gen{\bar S}_0A\supup{MHV}_{n}=0$ \cite{Witten:2003nn}.

In $(3,1)$ spacetime signature, however, $\lambda$ and $\bar \lambda$ are 
related by complex conjugation, and thus 
there is the holomorphic anomaly 
\cite{Cachazo:2004by,Cachazo:2004dr,Britto:2004nj}.
It gives a non-trivial contribution 
when the derivative with respect to $\bar\lambda$ acts on 
poles in the variable $\lambda$ (see \Appref{app:Conventions}). 
This gives rise to terms%
\footnote{We thank Emery Sokatchev for reminding us
of the precise form of the anomaly.}
\[\label{eq:spinoranomaly}
\frac{\partial}{\partial \bar{\lambda}^{\dot a} }\,
\frac{1}{\tprod{\lambda}{\mu}}
=\pi \deltad{2}(\tprod{\lambda}{\mu})\, \varepsilon_{\dot a \dot b}{\bar\mu}^{\dot b}.
\]
As can be immediately seen these anomaly terms coincide 
with the collinear singularities of the amplitude 
which as discussed previously is their physical origin. 

Here there is a crucial difference between the
$(3,1)$ physical Minkowski signature and 
$(2,2)$ split signature used for considerations in twistor space:
In $(3,1)$ signature two light-like vectors are orthogonal 
if and only if they are collinear. Collinearity implies two
constraints on the six degrees of freedom for two light-like vectors,
it is thus a codimension-two condition. 
In the spinor formulation collinearity is equivalent to 
$\tprod{\lambda_{k}}{\lambda_{k+1}}=0$ which is one complex
or two real conditions, hence codimension two.
Conversely for $(2,2)$ signature $\tprod{\lambda_{k}}{\lambda_{k+1}}=0$
is merely one real condition or codimension one.
Equivalently orthogonality does not imply full collinearity, but only 
one constraint.
The nature of the two types of singularities is rather different. 
The holomorphic anomaly only applies to codimension-two singularities. 
Codimension-one singularities can also have anomalies,
but one has to define properly the distributional meaning of
$(\tprod{\lambda_{k}}{\lambda_{k+1}})^{-1}$. 
One could consider adding $\pm i\epsilon$ to the denominators, 
but it is not clear which sign to use (for each term).
A principal value prescription appears to be the proper choice, 
but this leads to no anomaly. Altogether this consideration shows 
that the signature plays an important role for scattering amplitudes 
and we shall continue to work exclusively in Minkowski signature.

In the light of the holomorphic anomaly, there are extra terms 
in the action of $\gen{\bar S}_0$
\footnote{This fact and several of its implications discussed below 
have been found independently by James Drummond.
Johannes Henn and Emery Sokatchev also pointed 
out the distributional non-invariance.}
\<\label{eq:sbarona}
(\gen{\bar S}_0)^B_{\dot a}A\supup{MHV}_n
\eq
\sum_{k=1}^n\eta^B_k\,\frac{\partial}{\partial\bar\lambda^{\dot a}_{k}}\,
\frac{\deltad{4}(P)\,\deltad{8}(Q)}{\tprod{1}{2}\ldots\tprod{k-1}{k}\tprod{k}{k+1}\ldots \tprod{n}{1}}
\nln
\eq
-\pi\sum_{k=1}^n
\varepsilon_{\dot a \dot b}\bigbrk{\bar{\lambda}_{k-1}^{\dot b}\eta^B_k-\bar{\lambda}_{k}^{\dot b}\eta^B_{k-1}}
\frac{\deltad{2}(\tprod{\lambda_{k-1}}{\lambda_k})\,\deltad{4}(P)\,\deltad{8}(Q)}{\tprod{1}{2}\ldots\tprod{k-1}{k}^0\ldots \tprod{n}{1}}\,.
\>
The existence of extra terms is well-known 
and it has been employed successfully at the loop level 
\cite{Cachazo:2004by,Cachazo:2004dr,Britto:2004nj}.
At tree level, it has largely been ignored so far because the anomaly is restricted
to singular momentum configurations.

It turns out to be convenient to cast this statement 
into the language of generating functionals. 
Let $\mathcal{A}\supup{MHV}_n[J]$ be the generating functional
of MHV amplitudes \eqref{eq:MHV} with $n$ legs in the sense of \eqref{eq:AmpGen}.
Acting with the bare generator $\gen{\bar S}_0$
as defined in \eqref{eq:genfunc}
on $\mathcal{A}\supup{MHV}_n[J]$ 
by performing the functional variations and
integrating by parts we find
\<
\label{eq:sbarongena}
(\bar{\gen  S}_0)^{B}_{\dot a}{\cal A}\supup{MHV}_n[J]\eq
  -\pi 
   \int \prod_{k=1}^{n} d^{4|4}\Lambda_k \,
   \Tr\bigbrk{\comm{J(\Lambda_{1})}{ J(\Lambda_2)} \dots J(\Lambda_{n})}
\nl\qquad\qquad
  \times      \varepsilon_{\dot a \dot c}  \bar{\lambda}^{\dot c}_{1} \eta^B_2 \,
 \frac{\deltad{2}(\tprod{1}{2})\,\deltad{4}(P)\,\deltad{8}(Q)}{\tprod{2}{3}\dots\tprod{n}{1}}\,.
\>
We have made use of the cyclicity of the amplitudes, the trace and the measure 
in order to collect $n$ equivalent copies of the contribution in \eqref{eq:sbarona}
which thus cancel the symmetry factor of $1/n$ in \eqref{eq:AmpGen}.
The commutator term in the trace results from the 
difference term in the second line of \eqref{eq:sbarona} after interchanging
$\Lambda_1$ and $\Lambda_2$.

We can partially perform the integrals over $\Lambda_1$ to remove the delta function 
imposing the collinearity of the $1$ and $2$ legs. 
A convenient change of the variables $\Lambda_1$, $\Lambda_2$ to this end reads
\[\label{eq:varchange}
\begin{array}[b]{rclcrcl}
\lambda_1\eq e^{i \varphi} \lambda_{12} \sin \alpha,
&\quad&
\eta_1\eq e^{-i\varphi}(\eta_{12}\sin\alpha+\eta'\cos\alpha),
\\[1ex]
\lambda_2\eq \lambda_{12} \cos \alpha+z\lambda',
&\quad&
\eta_2\eq \eta_{12}\cos\alpha-\eta'\sin \alpha.
\end{array}
\]
The four complex variables $\lambda^a_1,\lambda^a_2$ have
been replaced by three complex variables $\lambda^{a}_{12},z$ and 
two real variables $\alpha\in[0,\half\pi]$, $\varphi\in[0,2\pi]$.
The spinor $\lambda'$ is a constant reference spinor.
The integral over $d^2z$ localises at $z=\bar z=0$ 
and after evaluating the various Jacobians we get that 
\<
\label{eq:sbaranom}
( \bar{\gen  S}_0)^{B}_{\dot a}\mathcal{A}\supup{MHV}_n[J]\eq
 -\pi  \int \prod_{k=3}^{n}  d^{4|4}\Lambda_k \, 
d^{4|4}\Lambda_{12}\, d^4 \eta'\, d\alpha  \, d\varphi
\,\, e^{3i\varphi} 
 \eps_{\dot a \dot c}  \bar{\lambda}^{\dot c}_1 \eta^B_2
  \times  
\nl\qquad\times
\Tr\bigbrk{\comm{ J(\Lambda_1)}{ J(\Lambda_2)}\dots J(\Lambda_{n})}
  \frac{\deltad{4}(P')\deltad{8}(Q')}{\tprod{12}{3}\dots\tprod{n}{12}} 
 \>
where  $P'=\lambda_{12}\bar{\lambda}_{12}+\sum_{k=3}^{n}\lambda_k\bar{\lambda}_k$ and
$Q'=\eta_{12} \lambda_{12}+\sum_{k=3}^{n}\eta_k \lambda_k$.  
Alternatively one can use the formula \eqref{eq:deltabracketeval} 
to derive this result.
Note that the integral over $\varphi$ amounts to the projection $\hat J(\Lambda_1)$ in \eqref{eq:Jhat}.
Removing the phase in $\Lambda_1$ such that 
$\lambda_1=\lambda_{12} \sin \alpha$ and $\eta_1=\eta_{12}\sin\alpha+\eta'\cos\alpha$
we obtain the more compact expression
\<
\label{eq:sbaranom2}
( \bar{\gen  S}_0)^{B}_{\dot a}\mathcal{A}\supup{MHV}_n[J]\eq
-2\pi^2 \int \prod_{k=3}^{n}  d^{4|4}\Lambda_k \, 
d^{4|4}\Lambda_{12}\, d^4 \eta'\, d\alpha   
\,\, \eps_{\dot a \dot c}  \bar{\lambda}^{\dot c}_1 \eta^B_2
  \times  
\nl\qquad\times
\Tr\bigbrk{\comm{ \hat J(\Lambda_1)}{ \hat J(\Lambda_2)}\dots J(\Lambda_{n})}
  \frac{\deltad{4}(P')\deltad{8}(Q')}{\tprod{12}{3}\dots\tprod{n}{12}} \,.
 \>
Note that the integrand is homogeneous in $\Lambda_2$ 
\eqref{eq:PhiHom} and 
thus the second source term $J(\Lambda_2)$ was replaced
by the projection $\hat J(\Lambda_2)$.

We observe that the anomalous variation \eqref{eq:sbaranom} produces $A\supup{MHV}_{n-1}$ 
with slight modifications merely on the first leg. 
Such a modification can be imposed through a variation 
of the sort $\int \Tr JJ\check J$ acting on $\mathcal{A}\supup{MHV}$.
More precisely the form of the correction
$\bar{\gen{S}}_+=\gen{\bar{S}}^{(0)}_{1,2}$ reads
\<\label{eq:Splus}
(\bar{\gen{S}}_+)^B_{\dot a}\eq
2\pi^2\int d^{4|4}\Lambda\,d^4 \eta'\,d\alpha\, \eps_{\dot a \dot c}  \bar{\lambda}_1^{\dot c} \, \eta_2^B 
\Tr\bigbrk{\comm{ \hat J(\Lambda_1)}{\hat J(\Lambda_2)} \check J(\Lambda)}
\nln\eq
-\pi^2\int d^{4|4}\Lambda\,d^4 \eta'\,d\alpha\, \eps_{\dot a \dot c} 
\bar{\lambda}^{\dot c} \eta'^B
\Tr\bigbrk{\comm{\hat J(\Lambda_1)}{\hat J(\Lambda_2)} \check J(\Lambda)}
\>
with the following definitions for $\Lambda_1$, $\Lambda_2$
\[\label{eq:Lambda12}
\begin{array}[b]{rclcrcl}
\lambda_1\eq \lambda \sin \alpha,
&\quad&
\eta_1\eq \eta\sin\alpha+\eta'\cos\alpha,
\\[1ex]
\lambda_2\eq \lambda \cos \alpha,
&\quad&
\eta_2\eq \eta\cos\alpha-\eta'\sin \alpha.
\end{array}
\]
The second form in \eqref{eq:Splus} is due to replacement of $\Lambda_1,\Lambda_2$
and making use of antisymmetry of the commutator.
The plus in $\bar{\gen{S}}_+$ signifies that the operator increases
the helicity by $+2$ relative to $\bar{\gen{S}}_0$. 
It was constructed such that
\[
\bar{\gen{S}}_0\mathcal{A}\supup{MHV}_n
+\bar{\gen{S}}_+\mathcal{A}\supup{MHV}_{n-1}=0.
\]

As can be seen we find a recursive pattern for the action of the generator
on the amplitudes. We can ask what is the starting point for this action and the 
answer is straightforward:
\[
\bar{\gen{S}}_0\mathcal{A}\supup{MHV}_4=0.
\]
This follows from the above calculation as
\[
(\gen{\bar S}_0)^B_{\dot a}A\supup{MHV}_4
=
-\pi\sum_{k=1}^4
\varepsilon_{\dot a \dot b}\bigbrk{\bar{\lambda}_{k-1}^{\dot b}\eta^B_k-\bar{\lambda}_{k}^{\dot b}\eta^B_{k-1}}
\frac{\deltad{2}(\tprod{\lambda_{k-1}}{\lambda_k})\,\deltad{4}(P)\,\deltad{8}(Q)}{\tprods{1}{2}\ldots\tprod{k-1}{k}^0\ldots \tprods{4}{1}}\,.
\]
However now, after making use of the delta-function imposing collinearity between
$p_k$ and $p_{k-1}$, the momentum conservation  implies that the three remaining 
momenta are collinear and the zero coming from the $\deltad{8}(Q)$ results in the
right hand side being zero. This is essentially equivalent to the fact that in $(3,1)$ signature
and for real momenta the three-point amplitude vanishes due to zero allowed phase space. 

In conclusion, the corrected classical superconformal generator 
(relevant to MHV amplitudes) is 
\[\label{eq:fullsbar}
\bar{\gen{S}}^B_{\dot a}=
(\bar{\gen{S}}_0)^B_{\dot a}+(\bar{\gen{S}}_+)^B_{\dot a}
\]
and it exactly annihilates the MHV functional $\mathcal{A}\supup{MHV}[J]$
\[
\bar{\gen{S}}\mathcal{A}\supup{MHV}[J]=0.
\]
Note that the cancellation is not restricted to the planar, 
but it holds for all $N\indup{c}$ and even for generic gauge groups.

Although, and as we will show later,  
it can be determined from the algebra, it is perhaps 
worthwhile to directly calculate $\gen{K}_+=\gen{K}^{(0)}_{1,2}$
from $\gen{K}_0$ acting on MHV amplitudes;
by a very similar calculation we find
\[\label{eq:Kplus}
(\gen{K}_+)_{b\dot a}=-2\pi^2  \int d^{4|4}\Lambda \, d^4\eta'\, d\alpha\,
\eps_{\dot a \dot c}  \bar{\lambda}_1^{\dot c} \Tr
\bigbrk{\comm{ \hat J( \Lambda_1 )}{\partial_{2,b} \hat J( \Lambda_2)} \check J(\Lambda)}
\]
with the same definitions as above.

\subsection{Conjugate MHV Amplitudes}
\label{sec:MHVbar}
Now we wish to find the deformation of the operator $\gen{S}$ defined in \eqref{eq:genfunc}
and the simplest method  is to  consider its action on $\overline{\mathrm{MHV}}$ amplitudes. 
A convenient form of the  tree-level $\overline{\mathrm{MHV}}$ contribution to the $n$-point super-amplitude 
is given by \cite{Drummond:2008bq}
\[
A\supup{\overline{MHV}}_n
=\deltad{4}(P)\,\deltad{8}(Q) F_{n}(\Lambda)
\]
where
\[
\deltad{8}(Q) F_{n}(\Lambda)=
\int \prod^n_{i=1} \bigbrk{d^4{\bar \eta}_i\, \exp(\eta^A\bar\eta_{i,A})}\, 
\frac{\deltad{8}({\bar Q})}{\ctprods{1}{2}\ctprods{2}{3}\ldots\ctprods{n}{1}}\,,
\qquad
\bar Q^{\dot a}_B=\sum_{i=1}^n\bar{\lambda}^{\dot a}_i {\bar \eta}_{i,B}.
\]
One can use the integral representation for the Gra{\ss}mannian delta function  
\[
\deltad{8}({\bar Q})=\int d^8 \omega\,  \exp\lrbrk{\omega^B_{\dot a}\bar Q^{\dot a}_B}
\]
to write this as 
\[
\deltad{8}(Q)F_{n}(\Lambda)=
\frac{1}{\ctprods{1}{2}\dots\ctprods{n}{1}}
\int d^8\omega \prod_{i=1}^n\deltad{4}(\eta_i-\bar{\lambda}^{\dot a}_i \omega_{\dot a}).
\]
Thus
\<
(\gen{S}_0)_{Ba}\mathcal{A}\supup{\overline{MHV}}[J]
\eq
\pi \sum_{n=4}^\infty  \int d^8\omega\prod_{k=1}^n 
\bigbrk{d^{4|4}\Lambda_k\,\deltad{4}(\eta_i-\bar{\lambda}_i^{\dot a} \omega_{\dot a})}
\frac{\deltad{2}(\ctprod{1}{2})\,\deltad{4}(P)}{\ctprod{2}{3}\dots\ctprod{n}{1}}
    \times
\nl\qquad
\times\eps_{ab}\lambda_1^b    
     \Tr\bigbrk{\comm{J(\Lambda_1)}{\partial_{2,B} J(\Lambda_2)}\dots J(\Lambda_{n})}.
\>
As was previously done we can use the delta function to partially perform the $\Lambda_n$
integral and rewrite the above expression
in a form that makes the necessary generator correction apparent. After a little algebraic
manipulation one can show that the correction
$\gen{S}_-=\gen{S}^{(0)}_{1,2}$ of $\gen{S}$ takes the form
\<\label{eq:s1+}
(\gen{S}_-)_{Ba}\eq 
-2\pi^2 \int d^{4|4}\Lambda \, d^4\eta'\,d\alpha \,\,
\deltad{4}(\eta')
\eps_{a  c}  {\lambda}_1^{c} \Tr
\bigbrk{\comm{\hat J(\Lambda_1)}{\partial_{2,B} \hat J(\Lambda_2)} \check J(\Lambda)}
\nln\eq
-2\pi^2 \int d^{4|4}\Lambda \, d\alpha \,\,
\eps_{a  c}  {\lambda}_1^{c} \Tr
\bigbrk{\comm{\hat J(\Lambda_1)}{\partial_{2,B}\hat J(\Lambda_2)} \check J(\Lambda)}
\nln\eq
+\pi^2 \int d^{4|4}\Lambda \, d^4\eta'\,d\alpha \,\,
\deltad{4}(\eta')
\eps_{a  c}  {\lambda}^{c}\partial'_{B}  \Tr
\bigbrk{\comm{\hat J(\Lambda_1)}{\hat J(\Lambda_2)} \check J(\Lambda)}
\>
where the minus in $\gen{S}_-$ signifies a decrease of the helicity.
As before $\Lambda_1$ and $\Lambda_2$ are defined in 
\eqref{eq:Lambda12} but with $\eta'=0$ for the second line.
The complete classical expression for the superconformal generator
\[\label{eq:fulls}
\gen{S}_{Ba}=
(\gen{S}_0)_{Ba}+(\gen{S}_-)_{Ba}
\]
annihilates the $\overline{\mathrm{MHV}}$ amplitudes
$\gen{S}\mathcal{A}\supup{\overline{MHV}}[J]=0$.

Similarly we find
\begin{equation}\label{eq:Kminus}
 (\gen K_-)_{a\dot a}=-2\pi^2 \int \dd^{4|4}\Lambda\, \dd^4\eta'\,\dd\alpha\,\, 
\deltad{4}(\eta')\Tr \bigcomm{\varepsilon_{ a  b}\lambda^{ b}_1\hat J(\Lambda_1)}{\bar\partial_{2,\dot a}\hat J(\Lambda_2)}\check J(\Lambda).
\end{equation}

\subsection{Negative-Energy Particles}
\label{sec:NegEn}

In the above discussion we have restricted ourselves to external
particles with positive energy, that is
$p_k^{a\dot{b}}=\lambda_k^a\tilde\lambda_k^{\dot{b}}$ with
$\tilde\lambda_k=+\bar\lambda_k$ and thus
$E_k=p_k^0=\half(\lambda_k^1\bar\lambda_k^1+\lambda_k^2\bar\lambda_k^2)>0$
for all particles $k$. As mentioned previously though, physical
scattering amplitudes require that at least two external particles
have negative energy, i.e.\ $\tilde\lambda_k=-\bar\lambda_k$ such that
$p_k^{a\dot{b}}=-\lambda_k^a\bar\lambda_k^{\dot{b}}$ and hence
$E_k<0$. In the following
we will extend our framework slightly in order to include
negative-energy particles.

In the tree-level MHV amplitude \eqref{eq:MHV}, negative-energy
particles only introduce a change of sign inside the momentum
delta-function $\deltad{4}(P)$. The form of collinear singularities
as in \eqref{eq:sbarona} is therefore not affected by the energy signs
of the adjacent collinear particles. As we shall see below, only the
splitting of the collective momentum into two collinear pieces as in
\eqref{eq:Lambda12} changes slightly when the adjacent particles have
different energy signs.

For including particles with positive and negative energies into our
framework, we introduce two types of source fields: $J^+(\Lambda)$
corresponds to positive-energy particles while $J^-(\Lambda)$
corresponds to negative-energy ones. The amplitude generating
functional \eqref{eq:AmpGen} comprising all possible particle
configurations then becomes%
\footnote{An alternative to deal with negative-energy particles is 
to represent them through variations $\check J(\Lambda)$
where the sources $J(\Lambda)$ correspond to positive-energy particles only.
The amplitude would thus be promoted to a variational operator. 
This picture is equivalent to the canonical quantum field theory 
framework where the S-matrix is an operator acting on the Fock space.}
\begin{equation}
\mathcal{A}[J]
=\sum_{n=4}^\infty\frac{1}{n}
 \int\prod_{k=1}^n d^{4|4}\Lambda_k
 \sum_{\substack{s_j=\pm\\(1\leq j\leq n)}}
 \Tr\bigbrk{J^{s_1}(\Lambda_1)\ldots J^{s_n}(\Lambda_n)}\,
 A_n(\Lambda_1^{s_1},\ldots,\Lambda_n^{s_n})\,,
\label{eq:AmpGenNeg}
\end{equation}
where $A_n(\Lambda_1^{s_1},\ldots,\Lambda_n^{s_n})$ equals the amplitude $A_n(\Lambda_1,\ldots,\Lambda_n)$ with
$\bar\lambda_k$ and $\eta_k$ replaced by $s_k\bar\lambda_k$ and
$s_k\eta_k$.%
\footnote{Changing also the sign of the fermionic variable $\eta$ is
purely conventional.}

When extending the formalism in this way, also the free symmetry
generators \eqref{eq:genfunc} need to include variations with respect
to particles of both energy signs. The generator $\gen{\bar{S}}_0$ for
example becomes
\begin{equation}
\brk{\gen{\bar{S}}_0}_{\dot{a}}^B
=-\int d^{4|4}\Lambda\,\eta^B\tr\sum_{s=\pm}\bigbrk{\bar{\partial}_{\dot{a}}J^s(\Lambda)}\check J^s(\Lambda)\,.
\label{eq:sbar0neg}
\end{equation}
We will now calculate the classical non-linear correction to this generator.
This generalises the treatment in \Secref{sec:MHV} and will result in
correction terms $\gen{\bar{S}}^{(0)}_{s_0\to\lrbrc{s_1,s_2}}$
that split a leg with sign $s_0$ into two collinear particles with
signs $s_1$, $s_2$.
Acting with $\gen{\bar{S}}_0$ on $\mathcal{A}\supup{MHV}[J]$ is
completely analogous to \eqref{eq:sbarongena} and yields
\begin{multline}
\brk{\gen{\bar{S}}_0}_{\dot{a}}^B\mathcal{A}\supup{MHV}[J]
=-\pi\sum_{n=4}^\infty\int\prod_{k=1}^n d^{4|4}\Lambda_k\,
 \varepsilon_{\dot{a}\dot{c}}\bar{\lambda}_1^{\dot{c}}\eta_2^B
 \frac{\deltad{2}(\tprod{1}{2})\deltad{8}(Q)}{\tprod{2}{3}\cdots\tprod{n}{1}}\\\times
 \sum_{s_j=\pm}\deltad{4}(P)
 \tr\bigbrk{\comm{J_1^{s_1}}{J_2^{s_2}}J_3^{s_3}\cdots J_{n}^{s_{n}}}\,,
\label{eq:sbar0onMHVneg}
\end{multline}
where $P=\sum_{j=1}^np_j=\sum_{j=1}^ns_j\lambda_j\bar\lambda_j$ and
$Q=\sum_{j=1}^ns_j\lambda_j\eta_j$. Now
the terms in which $s_1=s_2$, i.e.\ the part where the collinear
particles have both positive or both negative energy is compensated by
a non-linear correction
$\gen{\bar{S}}_+^==\gen{\bar{S}}^{(0)}_{s\to\lrbrc{s,s}}$ which looks exactly as
\eqref{eq:Splus} with $J$ replaced by $J^s$ and including a sum over
$s=\pm$.

The terms of \eqref{eq:sbar0onMHVneg} in which the two collinear particles $1$ and $2$ have
different energy signs we split into a part with
$\abs{E_1}<\abs{E_2}$ and a part with
$\abs{E_1}>\abs{E_2}$. 
In the former part, the momentum
$\lambda_2\bar\lambda_2-\lambda_1\bar\lambda_1$ carries the sign $s_2$
of particle $2$, in the latter part it carries the opposite sign $s_1$. 
Using the fact that
$\deltad{2}(\tprods{1}{2})/\tprods{2}{3}\cdots\tprod{n}{1}$ is invariant under
an exchange of the labels $1$ and $2$, we can exchange those labels
in the latter part. It then combines with the former to
\begin{multline}
-\pi\sum_{n=4}^\infty\int_{\abs{E_1}<\abs{E_2}}\prod_{k=1}^n d^{4|4}\Lambda_k\,
\varepsilon_{\dot{a}\dot{c}}\frac{\deltad{2}(\tprod{1}{2})\deltad{8}(Q)}{\tprod{2}{3}\cdots\tprod{n}{1}}\\\times
\sum_{\substack{s_j=\pm\\2\leq j\leq n}}
\deltad{4}(P)\bigbrk{\bar\lambda_1^{\dot{c}}\eta_2^B-\bar\lambda_2^{\dot{c}}\eta_1^B}
\tr\bigbrk{\comm{J_1^{-s_2}}{J_2^{s_2}}J_3^{s_3}\cdots J_{n}^{s_{n}}}\,,
\label{eq:posnegterm}
\end{multline}
where
$P=s_2\brk{\lambda_2\bar\lambda_2-\lambda_1\bar\lambda_1}+\sum_{j=3}^{n}s_j\lambda_j\bar\lambda_j$.
As before we can now use the delta function
$\deltad{2}(\tprods{1}{2})$ to partially perform the $\lambda_1$ integral
by using \eqref{eq:deltatrans},
this time setting $\lambda_1=e^{i\varphi}\lambda_2\tanh\alpha$,
rescaling $\lambda_2\to\lambda_2'\cosh\alpha$
and integrating over $\varphi$ and $\alpha$ instead of $\lambda_1$.
Further including a rotation of $\eta_1$ and $\eta_2$, altogether we
define the new set of variables $\Lambda_2'$, $\eta'$, $\alpha$ and
$\varphi$ through (cf.\ \eqref{eq:varchange}):
\begin{gather}
\label{eq:vardef2}
\begin{aligned}[b]
\lambda_1 &= e^{i\varphi}\lambda_{12}\sinh\alpha\,, \qquad&
\eta_1    &= e^{-i\varphi}\brk{\eta_{12}\sinh\alpha+\eta'\cosh\alpha}\,, \\
\lambda_2 &= \lambda_{12}\cosh\alpha\,, &
\eta_2    &= \eta_{12}\cosh\alpha+\eta'\sinh\alpha\,,
\end{aligned}\\
\Rightarrow
d^4\lambda_1\,d^4\lambda_2\,\deltad{2}(\tprods{1}{2})
=d\lambda_{12}\,d\varphi\,d\alpha\,\sinh\alpha\cosh\alpha\,.\nn
\end{gather}
With this change of variables, the part of
$\brk{\gen{\bar{S}}_0}_{\dot{a}}^B\mathcal{A}\supup{MHV}[J]$ where
$s_1=-s_2$ \eqref{eq:posnegterm} becomes
\begin{multline}
-\sum_{n=4}^\infty\int\prod_{k=3}^{n}d^{4|4}\Lambda_k\,d^{4|4}\Lambda_{12}
\sum_{\substack{s_j=\pm\\1\leq j\leq n-1}}\frac{\deltad{4}(P')\deltad{4}(Q')}{\tprod{12}{3}\cdots\tprod{n}{12}}
\cdot\pi\int d^4\eta'd\varphi\,d\alpha\,e^{3i\varphi}\\\times
\varepsilon_{\dot{a}\dot{c}}\brk{\bar\lambda_1^{\dot{c}}\eta_2^B-\bar\lambda_2^{\dot{c}}\eta_1^B}
\tr\bigbrk{\comm{J^{-s_2}_1}{J^{s_2}_2}J_3^{s_3}\cdots J_{n}^{s_{n}}}\,,
\label{eq:sbar0onMHVneg2}
\end{multline}
where $P'=s_2\lambda_{12}\bar\lambda_{12}+\sum_{j=3}^{n}s_j\lambda_j\bar\lambda_j$
and $Q'=s_2\lambda_{12}\eta_{12}+\sum_{j=3}^{n}s_j\lambda_j\eta_j$. As in the
purely positive-energy case \eqref{eq:sbaranom}, this produces
something very reminiscent of $A\supup{MHV}_{n-1}$ and can hence be
compensated by adding a term
$\gen{\bar{S}}^{(0)}_{s\to\lrbrc{+,-}}$ to $\gen{\bar{S}}$. In this case,
the correction term splits a particle with sign $s$ into two collinear
particles with opposite energy signs.

The complete tree-level correction to the operator
$\gen{\bar{S}}$ thus reads
\begin{equation}
\gen{\bar{S}}_+
=\gen{\bar{S}}^{(0)}_{s\to\lrbrc{s,s}}+\gen{\bar{S}}^{(0)}_{s\to\lrbrc{+,-}}
=\gen{\bar{S}}_+^=+\gen{\bar{S}}_+^{\neq}\,,
\label{eq:sbar1complete}
\end{equation}
where $\gen{\bar{S}}_+^=$ is given by \eqref{eq:Splus} with $J$
replaced by $J^s$ and including a sum over $s=\pm$. After replacing
$J^s$ by the projection $\hat J^s$ \eqref{eq:Jhat} and removing the
phase of $\Lambda_n$ in \eqref{eq:sbar0onMHVneg2}, the further
correction $\gen{\bar{S}}_+^{\neq}$ is given by%
\footnote{Note that the integral over $\alpha$ in \eqref{eq:sbar1pm}
runs from $0$ to $\infty$, while it runs from $0$ to $\half\pi$ in
\eqref{eq:Splus}.}
\begin{equation}
\brk{\gen{\bar{S}}_+^{\neq}}_{\dot{a}}^B
=2\pi^2\int d^{4|4}\Lambda\,d^4\eta'd\alpha\sum_{s=\pm}
 \varepsilon_{\dot{a}\dot{c}}\brk{\bar\lambda_1^{\dot{c}}\eta_2^{B}-\bar\lambda_2^{\dot{c}}\eta_1^{B}}
 \tr\bigbrk{\comm{\hat J^{-s}(\Lambda_1)}{\hat J^s(\Lambda_2)}\check J^s(\Lambda)}\,,
\label{eq:sbar1pm}
\end{equation}
where $\Lambda_1$ and $\Lambda_2$ are defined as
\begin{alignat}{2}
\lambda_1 &= \lambda\sinh\alpha\,, &\qquad
\eta_1    &= \eta\sinh\alpha+\eta'\cosh\alpha\,,\nn\\
\lambda_2 &= \lambda\cosh\alpha\,, &
\eta_2    &= \eta\cosh\alpha+\eta'\sinh\alpha\,.
\label{eq:varchangepm}
\end{alignat}

As can be seen from this example calculation, the contributions to the
classical generators coming from the inclusion of negative-energy
particles are obtained straightforwardly once the purely
positive-energy corrections are known. Since the additional terms
obscure notation though, we refrain from including them in the
remainder of this work.

\section{Closure of the Algebra}
\label{sec:Closure}

\begin{figure}
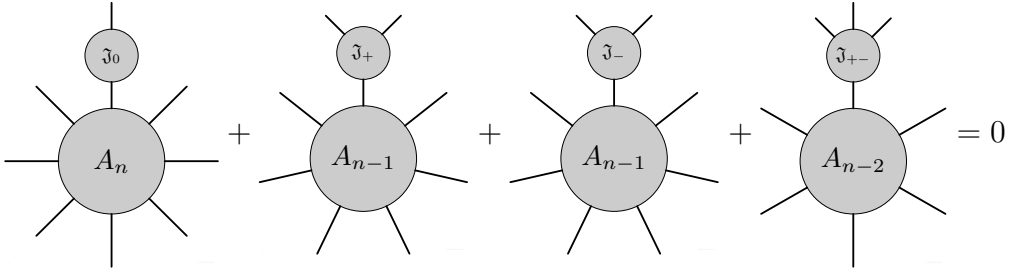
\centering
$
\includegraphicsbox[scale=0.7]{FigAmpTreeJ0.mps}
+\includegraphicsbox[scale=0.7]{FigAmpTreeJp.mps}
+\includegraphicsbox[scale=0.7]{FigAmpTreeJm.mps}
+\includegraphicsbox[scale=0.7]{FigAmpTreeJpm.mps}
=0$
\caption{Statement of exact invariance of tree amplitudes 
under the deformed superconformal representation.}
\label{fig:DynamicInvarianceTree}
\end{figure}

In the previous section perturbative corrections 
to the superconformal generators $\gen S_{aA}$ and $\bar{\gen S}_{\dot a}^B$ 
of $\superN=4$ SYM theory were derived by requiring the generating functional 
of MHV scattering amplitudes \eqref{eq:AmpGen} to be invariant 
under the action of these operators (cf.\ \Figref{fig:DynamicInvarianceTree}).

A priori, however, it is not clear that these deformations
are complete because we have considered only a subset of amplitudes.
An indication of completeness may come from algebra. 
We would like to show that the deformed generators 
still obey the $\alg{psu}(2,2|4)$ superconformal algebra,
which is also not clear a priori.

\subsection{Classical Representation}

Looking ahead to \Secref{sec:K},
the corrected generators are of the form
(cf.\ \Figref{fig:NLin})
\begin{figure}\centering
$
\gen{J}=
\includegraphicsbox[scale=1]{FigNLin0.mps}
+\includegraphicsbox[scale=1]{FigNLinP.mps}
+\includegraphicsbox[scale=1]{FigNLinM.mps}
+\includegraphicsbox[scale=1]{FigNLinPM.mps}
$
\caption{The free superconformal generators $\gen J_0$ are deformed by contributions 
changing the number of particles and thereby relating scattering 
amplitudes with different numbers of legs to each other.}
\label{fig:NLin}
\end{figure}
\[
\gen{S}=\gen{S}_0+\gen{S}_-,\qquad
\bar{\gen{S}}=\bar{\gen{S}}_0+\bar{\gen{S}}_+,\qquad
\gen K=\gen K_0+ \gen K_++ \gen K_-+ \gen K_{+-}.
\]
All other generators remain undeformed.
The correction terms to \eqref{eq:genfunc} 
were computed in 
\eqref{eq:Splus,eq:s1+,eq:Kplus,eq:Kminus} 
and read
\begin{align}
(\gen{S}_-)_{Aa}&= 
-2\pi^2 \int d^{4|4}\Lambda \, d^4\eta'\,d\alpha \,\,
\deltad{4}(\eta')
\Tr\comm{\eps_{a  b}  {\lambda}_1^{b}\hat J_1}{\partial_{2,A} \hat J_2} \check J,
\nonumber\\
(\bar{\gen S}_+)^A_{\dot a}&=+2\pi^2\int  \dd^{4|4}\Lambda \,\dd^4\eta'\,\dd\alpha\, 
\Tr\bigcomm{\varepsilon_{\dot a\dot b}\bar\lambda_1^{\dot b}\hat J_1}{\eta_2^A\hat J_2}\check J,
\nonumber\\
(\gen K_-)_{a\dot a}&=-2\pi^2 \int \dd^{4|4}\Lambda\, \dd^4\eta'\,\dd\alpha\,\, 
\deltad{4}(\eta')\Tr \bigcomm{\varepsilon_{ a  b}\lambda^{ b}_1\hat J_1}{\bar\partial_{2,\dot a}\hat J_2}\check J,
\nonumber\\
(\gen K_+)_{a\dot a}&=-2\pi^2\int  \dd^{4|4}\Lambda \,\dd^4 \eta'\, \dd\alpha\,
 \Tr \bigcomm{\varepsilon_{\dot a\dot b}\bar\lambda_1^{\dot b}\hat J_1}{\partial_{2,a}\hat J_2}\check J,
\label{eq:defgen}
\end{align}
where $J_k=J(\Lambda_k)$, $J=J(\Lambda)$.
The term $\gen{K}_{+-}$ can be found at the end of \Secref{sec:K}.
The spinor helicity coordinates $\Lambda_1$, $\Lambda_2$ are defined as follows \eqref{eq:Lambda12}
\begin{align}
\lambda_1&= \lambda \sin \alpha,
&\eta_1&=  \eta \sin \alpha+  \eta'  \cos \alpha,
\nonumber\\
\lambda_2&=\lambda \cos  \alpha,
&\eta_2&=\eta \cos \alpha-\eta' \sin \alpha.
\end{align}

\subsection{Algebra Relations}

It is straight-forward to read off the algebra relations 
from the representation \eqref{eq:StdRep} of the undeformed generators. 
The indices of a generator $\gen J$ under Lorentz and internal symmetry 
transform as
\begin{align}
 [\gen L^a{}_b,\gen J_c]&=-\delta^a_c \gen J_b+\half \delta^a_b \gen J_c,
&[\gen L^a{}_b,\gen J^c]&=\delta^c_b \gen J^a-\half \delta^a_b \gen J^c,
\nonumber\\
[\gen R^A{}_B,\gen J_C]&=-\delta^A_C\gen J_B+\quarter \delta^A_B\gen J_C,
&[\gen R^A{}_B,\gen J^C]&=\delta^C_B\gen J^A-\quarter \delta^A_B\gen J^C,
\nonumber\\
[\bar{\gen L}^{\dot a}{}_{\dot b},\gen J_{\dot c}]&=-\delta_{\dot c}^{\dot a}\gen J_{\dot b}+\half\delta^{\dot a}_{\dot b}\gen J_{\dot c},
&[\bar{\gen L}^{\dot a}{}_{\dot b},\gen J^{\dot c}]&=\delta^{\dot c}_{\dot b}\gen J^{\dot a}-\half\delta^{\dot a}_{\dot b}\gen J^{\dot c}.
\end{align}
All indices in the deformations \eqref{eq:defgen}
are contracted properly using only invariant symbols.
Consequently all commutators with $\gen L$, $\bar{\gen L}$ and $\gen R$
are unchanged using the free rotation generators
$\gen L_0$, $\bar{\gen L}_0$ and $\gen R_0$.

Commutators with the dilatation generator, $[\gen D,\gen J]=\Dim(\gen J) \gen J$, 
are specified through the conformal dimensions of the generators, the non-trivial ones being
\[
\mathrm{dim}(\gen P)=-\mathrm{dim}(\gen K)=1,\qquad \mathrm{dim}(\gen Q)=\mathrm{dim}(\bar{\gen Q})=-\mathrm{dim}(\gen S)=-\mathrm{dim}(\bar{\gen S})=\half.
\]
By power counting it is also straight-forward to show that $\gen{D}=\gen{D}_0$
yields the correct algebra.

It is the aim of this section to show that the additional non-trivial algebra relations given by
\begin{align}
\acomm{\gen Q^{aA}}{\bar{\gen Q}^{\dot a}_B}&=\delta^A_B\gen P^{a\dot a}, 
&\acomm{\gen S_{aA}}{\bar{\gen S}_{\dot a}^B}&=\delta^B_A \gen K_{a\dot a},
\nonumber\\
\comm{\gen P^{a\dot a}}{\gen S_{b A}}&=\delta^a_b\bar{\gen Q}^{\dot a}_A,
&\comm{\gen K_{a\dot a}}{\gen Q^{bA}}&=\delta_a^b\bar{\gen S}_{\dot a}^A,
\nonumber\\
\comm{\gen P^{a\dot a}}{\bar{\gen S}_{\dot b}^A}&=\delta^{\dot a}_{\dot b}{\gen Q}^{aA},
&\comm{\gen K_{a\dot a}}{\bar{\gen Q}^{\dot b}_A}&=\delta_{\dot a}^{\dot b}{\gen S}_{aA},
\end{align}
and 
\begin{align}
\comm{\gen K_{a\dot a}}{\gen P^{b\dot b}}&=\delta_{\dot a}^{\dot b}\gen L^b{}_a+\delta_a^b\bar{\gen L}^{\dot b}{}_{\dot a}+\delta_a^b\delta_{\dot a}^{\dot b}\gen D,
\nonumber\\
\acomm{\gen Q^{aA}}{\gen S_{bB}}&=\delta^A_B\gen L^a{}_b-\delta_b^a\gen R^A{}_B+ \delta_b^a\delta_B^A(\half\gen D+\quarter\gen C),
\nonumber\\
\acomm{\bar{\gen Q}^{\dot a}_A}{\bar{\gen S}_{\dot b}^B}&=\delta_A^B\bar{\gen L}^{\dot a}{}_{\dot b}+\delta_{\dot b}^{\dot a}\gen R^B{}_A+\delta_{\dot b}^{\dot a}\delta_A^B(\half\gen D-\quarter\gen C),
\end{align}
as well as all trivial commutators are not altered by the introduced corrections. 

Since $\gen P$ and $\gen K$ are expressed in terms of $\gen Q$, $\bar{\gen Q}$ 
and $\gen S$, $\bar{\gen S}$, respectively, the verification of the algebra 
reduces to a minimal set of commutation relations. 
These are the relations involving only the four latter operators. 
The remaining commutators then follow using the Jacobi identity 
as will be demonstrated at the end of the section.

\subsection{The Generator \texorpdfstring{$\gen{K}$}{K}}
\label{sec:K}

For a verification of the superconformal algebra it is not necessary 
to explicitly construct the generator $\gen K$ of special conformal transformations. 
Nevertheless it is desirable to obtain an expression for its deformations 
with regard to the symmetries of scattering amplitudes. 
The corrections to the conformal generator take the form 
\begin{align}
 \gen K&=\gen K_0+ \gen K_++ \gen K_-+ \gen K_{+-}
\nonumber\\
&=\acomm{\gen S_0}{\bar{\gen S}_0}
+\acomm{\gen S_0}{\bar{\gen S}_+}
+\acomm{\gen S_-}{\bar{\gen S}_0}
+\acomm{\gen S_-}{\bar{\gen S}_+}.
\end{align}
Employing the expressions for the corrections to $\gen S$ and $\bar{\gen S}$ obtained in \eqref{eq:Splus,eq:s1+},
the above anti-commutators can be explicitly evaluated. 
We make use of the notation introduced in \eqref{eq:Lambda12}
and note the following set of useful identities for the evaluation of commutation relations
\begin{align}
 0&=\lambda^a \eta^A- \lambda_1^a \eta_1^A- \lambda_2^a \eta_2^A,\label{eq:lambet}\\
\eta_1^B\bar\partial_{1,\dot a} J_1&=(\eta^B+\cot{\alpha} \, \eta'^B)\bar\partial_{\dot a}J_1,\label{eq:etadel1}\\
  \eta_2^B\bar\partial_{2,\dot a} J_2&=(\eta^B-\tan{\alpha} \, \eta'^B)\bar\partial_{\dot a}J_2,\label{eq:etadel2}\\
\bar\lambda_1^{\dot a}\partial_{1,A} J_1&=\bar\lambda^{\dot a}\partial_{A} J_1,\label{eq:lambdel1}\\
\bar\lambda_2^{\dot a}\partial_{2,A} J_2&=\bar\lambda^{\dot a}\partial_{A} J_2.\label{eq:lambdel2}
\end{align}

We first compute the anti-commutator of $\gen{\bar S}_0$ and $\gen{S}_-$ to find 
\begin{align}
\bigacomm{(\gen S_-)_{aA}}{(\bar{\gen S}_0)_{\dot a}^B}\hat J
&=2\pi^2 \int \dd^4\eta'\,\dd\alpha\, \deltad{4}(\eta')\,\varepsilon_{ a b}\lambda^{b}_1
\Big\{
-\eta^B\bar\partial_{\dot a}\bigcomm{\hat J_1}{\partial_{2,A} \hat J_2}
\nln
&\qquad\qquad
+\eta_1^B\bigcomm{\bar\partial_{1,\dot a}\hat J_1}{\partial_{2,A} \hat J_2}
-\bigcomm{\hat J_1}{\partial_{2,A} \eta_2^B\bar\partial_{2,\dot a}\hat J_2}
\Big\}.
\end{align}
Evaluating \eqref{eq:etadel1,eq:etadel2} 
at $\eta'=0$ yields
$\acomm{(\gen S_-)_{aA}}{(\bar{\gen S}_0)_{\dot a}^B}=\delta_A^B(\gen K_-)_{a\dot a}$
with
\[
(\gen K_-)_{a\dot a}\hat J=-2\pi^2 \int \dd^4\eta'\,\dd\alpha\, \deltad{4}(\eta')\,
\bigcomm{\varepsilon_{ a  b}\lambda^{ b}_1 \hat J_1}{\bar\partial_{2,\dot a} \hat J_2}.
\]

In order to compute $\gen K_+$ we consider
\begin{align}
\bigacomm{(\gen S_0)_{aB}}{(\bar {\gen S}_+)_{\dot a}^A}\hat J
&=2\pi^2 \int \dd^4\eta'\,\dd\alpha\, \varepsilon_{\dot a \dot b}\bar\lambda^{\dot b}_1
\Big\{
-\partial_{a}\partial_{B}\bigcomm{\hat J_1}{\etaP{2}^A \hat J_2}\nln
&\qquad\qquad
-\etaP{2}^A \bigcomm{\partial_{1,a}\partialP{1,B}\hat J_1}{\hat J_2}
-\etaP{2}^A \bigcomm{\hat J_1}{\partial_{2,a}\partialP{2,B}\hat J_2}
\Big\}.
\end{align}
We add the following integration by parts term to the r.h.s.
\[
2\pi^2 \int \dd^4\eta'\,\dd\alpha\,\partial'_{B}\, \varepsilon_{\dot a \dot b}\bar\lambda^{\dot b}_1
\Big\{
-\cos\alpha\,\bigcomm{\partial_{1,a}\hat J_1}{\etaP{2}^A \hat J_2}
+\sin\alpha\,\bigcomm{\hat J_1}{\etaP{2}^A \partial_{2,a}\hat J_2}
\Big\}=0
\]
in order to shift all fermionic derivatives to their bosonic counterparts. 
Using $\partial_{A}J_1=\sin\alpha\,\partial_{1,A}J_1$
and $\partial'_{A}J_1=\cos\alpha\,\partial_{1,A}J_1$, etc., we obtain
$\acomm{(\gen S_0)_{aB}}{(\bar {\gen S}_+)_{\dot a}^A}=\delta_B^A (\gen K_+)_{a\dot a}$ with
\[
(\gen K_+)_{a\dot a}\hat J=-2\pi^2 \int \dd^4\eta'\,\dd\alpha\, 
\bigcomm{\varepsilon_{\dot a \dot b}\bar\lambda^{\dot b}_1\hat J_1}{\partial_{2,a}\hat J_2},
\]
which coincides with the result given in \eqref{eq:Kplus}.

Finally, we want to show that $\acomm{\gen S_-}{\bar{\gen S}_+}$
is a $\alg{su}(4)$ singlet and hence defines $\gen K_{+-}$ properly.
To make the calculation more tractable, we introduce two sets of
fermionic variables $\tilde\theta^A,\theta^A$ which we contract with 
the generators
\[
\gen{S}^a_- := \varepsilon^{ab}\tilde\theta^{C}(\gen{S}_-)_{bC},
\qquad
\gen{\bar S}^{\dot a}_+ := \varepsilon^{\dot a\dot b}\varepsilon_{CDEF}\theta^{D}\theta^{E}\theta^{F}(\gen{\bar S}_+)^C_{\dot b}.
\]
The aim is to show that the commutator is
totally antisymmetric in $\tilde\theta$ and the $\theta$'s%
\footnote{The fermionic variables $\theta$ turn the new generators $\gen{S}^a_-$ and $\gen{\bar S}^{\dot a}_+$
into bosonic operators. Consequently we should compute their commutator. 
Likewise all the objects in the following computation will turn out to be (conveniently) bosonic.}
\[\label{eq:ssbaraim}
\bigcomm{\gen{S}^a_-}{\gen{\bar S}^{\dot b}_+} 
\sim \varepsilon_{CDEF}\tilde\theta^{C}\theta^{D}\theta^{E}\theta^{F}.
\]
We can evaluate the action of the generators on a source
by rewriting the fermionic integral in $\gen{\bar S}^{\dot a}_+$
as $\int d^4\eta'\,\eta'\sim \partial'^3$
\<\label{eq:ssbarexpand}
\gen{S}^a_- \hat J(\Lambda)
\earel{\sim}
+\lambda^a\int d\alpha\,\cos\alpha\,\bigcomm{\tilde\partial \hat J(\sin \alpha\Lambda)}{J(\cos\alpha\Lambda)}
\nl
-\lambda^a\int d\alpha\,\sin\alpha\,\bigcomm{\hat J(\sin \alpha\Lambda)}{\tilde\partial J(\cos\alpha\Lambda)},
\nln
\gen{\bar S}^{\dot a}_+ \hat J(\Lambda)
\earel{\sim}
+\bar\lambda^{\dot a}\int d\alpha\,\cos^3\alpha\,\bigcomm{\partial^3 \hat J(\sin \alpha\Lambda)}{J(\cos\alpha\Lambda)}
\nl
-3\bar\lambda^{\dot a}\int d\alpha\,\cos^2\alpha\sin\alpha\,\bigcomm{\partial^2 \hat J(\sin \alpha\Lambda)}{\partial J(\cos\alpha\Lambda)}
\nl
+3\bar\lambda^{\dot a}\int d\alpha\,\cos\alpha\sin^2\alpha\,\bigcomm{\partial\hat J(\sin \alpha\Lambda)}{\partial^2 J(\cos\alpha\Lambda)}
\nl
-\bar\lambda^{\dot a}\int d\alpha\,\sin^3\alpha\,\bigcomm{\hat J(\sin \alpha\Lambda)}{\partial^3 J(\cos\alpha\Lambda)}.
\>
The above index-free partial derivatives are defined as $\partial:=\theta^A\partial_A$ and 
$\tilde\partial:=\tilde\theta^A\partial_A$, and they are bosonic.
Applying the two generators to a source $J(\Lambda)$ results in 
three sources $J_{x,y,z}:=J(x\Lambda,y\Lambda,z\Lambda)$ with spherical coordinates and measure
\[
x=\sin\alpha\cos\beta,\quad
y=\sin\alpha\sin\beta,\quad
z=\cos\alpha,\qquad
\int d^2\Omega=\int d\alpha\,d\beta\,\sin\alpha.
\]
The benefit of these coordinates is that they are fully interchangeable
which allows for the Jacobi identity to be used easily.

Using this expression we can 
compute $\comm{\gen{S}^a_-}{\gen{\bar S}^{\dot b}_+}\hat J(\Lambda)$
and find 16 terms initially which can be grouped into 5 classes 
depending on how their derivatives are distributed.
Some terms have to be converted by means of a Jacobi identity and 
permuting the coordinates $x,y,z$ accordingly.
It is now a matter of patience and care to show that all the derivatives $\partial$ and $\tilde\partial$ appear
symmetrically and thus \eqref{eq:ssbaraim} holds.

There is however a slightly more convenient way 
to show the required property formally:
We note that the terms in \eqref{eq:ssbaraim}
follow a certain regular pattern. 
Let us therefore introduce some derivative operators $\partial_1,\partial_2$
acting on three sources $J_{x,y,z}$ according to 
\[\begin{array}{rclcrclcrcl}
\partial_1 J_x\eq\displaystyle\frac{xz}{\sqrt{1-z^2}}\,\partial J_x,
&\quad&
\partial_1 J_y\eq\displaystyle\frac{yz}{\sqrt{1-z^2}}\,\partial J_y,
&\quad&
\partial_1 J_z\eq\displaystyle -\sqrt{1-z^2}\,\partial J_z,
\\[1ex]
\partial_2 J_x\eq\displaystyle \frac{y}{\sqrt{1-z^2}}\,\partial J_x,
&\quad&
\partial_2 J_y\eq\displaystyle \frac{-x}{\sqrt{1-z^2}}\,\partial J_y,
&\quad&
\partial_2 J_z\eq\displaystyle 0,
\end{array}
\]
It is easy to convince oneself that 
\[
\label{eq:ssbartheta}
\bigcomm{\gen{S}^a_-}{\gen{\bar S}^{\dot b}_+}\hat J(\Lambda)
\sim 
\lambda^a\bar\lambda^{\dot b}
\int d^2\Omega
\lrbrk{
\tilde\partial_1(\partial_2)^3
-\tilde\partial_2(\partial_1)^3
}
\bigcomm{\comm{\hat J_x}{\hat J_y}}{\hat J_z}.
\]
The point is that $\gen{S}^a_-\sim \partial_k$
and $\gen{S}^{\dot b}_+\sim (\partial_k)^3$,
cf.\ \eqref{eq:ssbarexpand},
and the index $k$ tells whether the operator
acts on the outer or the inner commutator.
Note that the density factor $\sin\alpha$ of $d^2\Omega$ originates
from rescaling $\lambda^a$ or $\bar\lambda^{\dot a}$ in the second generator.

The above expression \eqref{eq:ssbartheta} is however not yet manifestly symmetric
in tilded and untilded derivatives as required for \eqref{eq:ssbaraim}.
We have to use the Jacobi identity to achieve symmetry. It turns out that
replacing
\[
\bigcomm{\comm{\hat J_x}{\hat J_y}}{\hat J_z}
\to
\sfrac{2}{3}\bigcomm{\comm{\hat J_x}{\hat J_y}}{\hat J_z}
-\sfrac{1}{3}\bigcomm{\comm{\hat J_x}{\hat J_z}}{\hat J_y}
-\sfrac{1}{3}\bigcomm{\comm{\hat J_z}{\hat J_y}}{\hat J_x}
\] 
achieves the goal.
In order to make the three terms comparable, we have to 
permute the coordinates $x,y,z$. The permutations
also transform the two derivative operators 
$\vec\partial=(\partial_1,\partial_2)$
using the permutation matrices
\<
P_{xy}\vec\partial\eq \matr{cc}{+1&0\\0&-1}\vec\partial,
\nln
P_{xz}\vec\partial\eq\frac{1}{\sqrt{1-x^2}\sqrt{1-z^2}}\matr{cc}{-xz&-y\\-y&+xz}\vec\partial,
\nln
P_{yz}\vec\partial\eq\frac{1}{\sqrt{1-y^2}\sqrt{1-z^2}}\matr{cc}{-yz&+x\\+x&+yz}\vec\partial.
\>
In confirming the relation one can for convenience treat 
$\partial_k$ as two bosonic variables and thus 
\eqref{eq:ssbartheta} is merely a quadratic polynomial in $\partial_k$.

Using the same notation we can formally write down $\gen{K}_{+-}$
\[\label{eq:Kplusminus}
(\gen{K}_{+-})_{b\dot a}
\sim 
\int d^{4|4}\Lambda\,d^2\Omega\,d^4\theta\,
\varepsilon_{bd}\varepsilon_{\dot a\dot c}\lambda^d\bar\lambda^{\dot c}
\lrbrk{\partial_1(\partial_2)^3-\partial_2(\partial_1)^3}
\bigcomm{\comm{\hat J_x}{\hat J_y}}{\hat J_z}\check J.
\]
Due to $\gen K\sim\acomm{\gen{S}}{\bar{\gen S}}$ the conformal generator 
inherits the property to annihilate the generating functional 
of scattering amplitudes \eqref{eq:AmpGen} from $\gen{S}$, $\bar{\gen S}$. 
We thus consider it an unreasonable hardship to compute the precise prefactor
of \eqref{eq:Kplusminus}.

\subsection{Commutators between \texorpdfstring{$\gen{Q}$}{Q}, \texorpdfstring{$\bar{\gen{Q}}$}{Q bar}
and \texorpdfstring{$\gen{S}$}{S}, \texorpdfstring{$\bar{\gen{S}}$}{S bar}}

In this section we demonstrate that the commutation relations 
of the generators $\gen S$, $\bar{\gen S}$ with $\gen{Q}$, $\bar{\gen{Q}}$ 
are not altered by the perturbative corrections introduced above by acting on 
a source term $\hat J(\Lambda)$. 

It is straight-forward to show that the anticommutator 
between $\gen Q_0$ and $\bar{\gen S}_+$ 
vanishes by means of \eqref{eq:lambet}
\begin{equation}\label{eq:comQ0S1bar}
 \bigacomm{(\gen Q_0)^{aA}}{(\bar{\gen S}_+)_{\dot a}^B}=0.
\end{equation}
Taking into account \eqref{eq:lambdel1,eq:lambdel2},
also the anti-commutator of $\bar{\gen Q}_0$ and $\gen S_-$ vanishes:
\begin{align}
\bigacomm{(\bar{\gen Q}_0)^{\dot a}_A}{(\gen S_-)_{aB}}\hat J&=
2\pi^2\int  \dd^4 \eta'\, \dd\alpha\, \deltad{4}(\eta')\,\varepsilon_{ab}\lambda_1^b
\Big\{\bar\lambda_1^{\dot a} \bigcomm{\partialP{1, A}\hat J_1}{\partialP{2,B}\hat J_2}\nln
&\qquad\qquad+\bar\lambda_2^{\dot a}\bigcomm{\hat J_1}{\partialP{2,A}\partialP{2,B}\hat J_2}
-\bar \lambda^{\dot a}\partial_A\bigcomm{\hat J_1}{\partialP{2,B}\hat J_2}\Big\}\nln
&=0.
\end{align}

Next we evaluate the anti-commutator of $\gen Q_0$ and $\gen S_-$ giving
\begin{align}
\bigacomm{({\gen Q}_0)^{aA}}{({\gen S}_-)_{bB}}\hat J&=
-2\pi^2 \int  \dd^4 \eta'\, \dd\alpha\, \deltad{4}(\eta')\varepsilon_{bc}\lambda_1^c
\Big\{\lambda_1^a\eta_1^A\bigcomm{\hat J_1}{\partial_{2,B}\hat J_2}\nln
&\qquad\qquad-\lambda_2^a\bigcomm{\hat J_1}{\partial_{2,B}(\eta_2^A\hat J_2)}
-\lambda^a\eta^A\bigcomm{\hat J_1}{\partial_{2,B}\hat J_2}\Big\}.
\end{align}
Now \eqref{eq:lambet} yields
\begin{equation}
\bigacomm{({\gen Q}_0)^{aA}}{({\gen S}_-)_{bB}}\hat J=
2\pi^2\delta_B^A \int  \dd^4 \eta'\, \dd\alpha\, \deltad{4}(\eta')\,
\varepsilon_{bc}\lambda_1^c\lambda_2^a\bigcomm{\hat J_1}{\hat J_2},
\end{equation}
and the integral is antisymmetric under the shift of the integration variable
\begin{equation}
 	\alpha\mapsto\frac{\pi}{2}-\alpha \quad\Rightarrow\quad \Lambda_1\leftrightarrow\Lambda_2,
\end{equation}
and does therefore vanish.

Last but not least we compute the anti-commutator of $\bar{\gen Q}_0$ and $\bar{\gen S}_+$ giving
\begin{align}
\bigacomm{(\bar{\gen Q}_0)_{A}^{\dot a}}{(\bar{\gen S}_+)_{\dot b}^B}\hat J&=
-2\pi^2 \int  \dd^4 \eta'\, \dd\alpha\, \varepsilon_{\dot b\dot c}\bar\lambda_1^{\dot c}
\Big \{-\bar\lambda^{\dot a}\partial_{A}\bigcomm{\hat J_1}{\etaP{2}^B\hat J_2}\nln
&\qquad\qquad-\etaP{2}^B\bar\lambda_1^{\dot a}\bigcomm{\partialP{1,A}\hat J_1}{\hat J_2}
-\etaP{2}^B\bar\lambda_2^{\dot a}\bigcomm{\hat J_1}{\partialP{2,A}\hat J_2} \Big\}.
\end{align}
By means of \eqref{eq:lambdel1,eq:lambdel2} 
we are left with an integral expression of the form
\begin{align}
\bigacomm{(\bar{\gen Q}_0)_{A}^{\dot a}}{(\bar{\gen S}_+)_{\dot b}^B}\hat J
&=2\pi^2\delta_A^B \int  \dd^4 \eta'\, \dd\alpha \,
\varepsilon_{\dot b\dot c}\bar\lambda_1^{\dot c}\bar\lambda_2^{\dot a}\bigcomm{\hat J_1}{\hat J_2}
=0.
\end{align}
The integral, however, again vanishes being antisymmetric under a shift of integration variables:
\begin{equation}
 	\alpha\mapsto\frac{\pi}{2}-\alpha,\qquad\eta'\mapsto -\eta'
\quad\Rightarrow\quad \Lambda_1\leftrightarrow\Lambda_2.
\label{eq:intshift}
\end{equation}

\subsection{Commutators between \texorpdfstring{$\gen{S}$}{S} and \texorpdfstring{$\bar{\gen{S}}$}{S bar}}

The anticommutator of two $\gen{S}$ vanishes in $\alg{psu}(2,2|4)$
and the same is true for the tree superconformal representation
$\gen{S}_0$; similarly for $\gen{\bar S}$.

Let us now compute the corrections due to $\gen{S}_-$ by acting
on the source $\hat J(\Lambda)$.
Straight-forward evaluation yields
\<
\acomm{({\gen S}_0)_{aB}}{({\gen S}_-)_{cD}}\hat J\eq
-2\pi^2\int \dd\alpha\,\varepsilon_{ac}\sin^2\alpha
\bigacomm{\partial_{1,B}\hat J_1}{\partial_{2,D}\hat J_2}
\nl
-2\pi^2\int \dd\alpha\,\varepsilon_{ce}\lambda^{e}\sin\alpha\cos^2\alpha
\bigacomm{\partial_{1,a}\partial_{1,B}\hat J_1}{\partial_{2,D}\hat J_2}
\nl
+2\pi^2 \int \dd\alpha\,\varepsilon_{ce}\lambda^{e}\sin^2\alpha\cos\alpha
\bigacomm{\partial_{1,B}\hat J_1}{\partial_{2,a}\partial_{2,D}\hat J_2}
\nl
-2\pi^2 \int \dd\alpha\,\varepsilon_{ac}\sin\alpha\cos\alpha
\bigcomm{\hat J_1}{\partial_{2,B}\partial_{2,D}\hat J_2}
\nl
-2\pi^2\int \dd\alpha\,\varepsilon_{ce}\lambda^{e}\sin^3\alpha
\bigcomm{\hat J_1}{\partial_{2,a}\partial_{2,B}\partial_{2,D}\hat J_2}
\nl
+2\pi^2 \int \dd\alpha\,\varepsilon_{ce}\lambda^{e}\sin^2\alpha\cos\alpha
\bigcomm{\partial_{1,a}\hat J_1}{\partial_{2,B}\partial_{2,D}\hat J_2}.
\>
The expansion of the anticommutator
$\acomm{{\gen S}_{aB}}{{\gen S}_{cD}}$ 
contains the above anticommutator
symmetrised over the pairs $aB$ and $cD$.
Note that each term in the above expression
is manifestly antisymmetric in $a,c$ or in $B,D$.
Thus the final term must be antisymmetric in
both $a,c$ and $B,D$. We can make the antisymmetry 
in $a,c$ manifest by pulling out $\varepsilon_{ac}$.
After flipping some of the integral regions, $\alpha\mapsto\half\pi-\alpha$, 
and rearranging some terms for later convenience,
we obtain for 
$\acomm{({\gen S}_0)_{aB}}{({\gen S}_-)_{cD}}
+\acomm{({\gen S}_0)_{cD}}{({\gen S}_-)_{aB}}$
\<\ldots
\eq
\pi^2\varepsilon_{ac}\int \dd\alpha\,(\cos^2\alpha-\sin^2\alpha)
\bigacomm{\partial_{1,B}\hat J_1}{\partial_{2,D}\hat J_2}
\nl
+\pi^2 \varepsilon_{ac}\int \dd\alpha\,\sin\alpha\cos\alpha
\bigacomm{\cot\alpha(2\lambda^{e}_1\partial_{1,e}+1)\partial_{1,B}\hat J_1}{\partial_{2,D}\hat J_2}
\nl
+\pi^2 \varepsilon_{ac}\int \dd\alpha\,\sin\alpha\cos\alpha
\bigacomm{\partial_{2,B}\hat J_1}{-\tan\alpha(2\lambda^{e}_2\partial_{2,e}+1)\partial_{2,D}\hat J_2}
\nl
-\pi^2 \varepsilon_{ac} \int \dd\alpha\,2\sin\alpha\cos\alpha
\bigcomm{\hat J_1}{\partial_{2,B}\partial_{2,D}\hat J_2}
\nl
-\pi^2 \varepsilon_{ac}\int \dd\alpha\,\sin^2\alpha
\bigcomm{\cot\alpha(2\lambda^{e}_1\partial_{1,e}+2)\hat J_1}{\partial_{2,B}\partial_{2,D}\hat J_2}
\nl
-\pi^2 \varepsilon_{ac}\int \dd\alpha\,\sin^2\alpha
\bigcomm{\hat J_1}{-2\tan\alpha\lambda^{e}_2\partial_{2,e}\partial_{2,B}\partial_{2,D}\hat J_2}
\>
We would like to recast all these integrands in
the form of a total derivative w.r.t.\ $\alpha$.
To this end we notice that terms like 
$\lambda^{e}\partial_{e}\hat J$
do appear in $\partial_\alpha \hat J$.
Conversely, contributions of the sort
$\bar\lambda^{\dot e}\bar\partial_{1,\dot e}$ and $\eta^{E}\partial_{1,E}$
which are also part of $\partial_\alpha \hat J_{1,2}$ to not appear. 
To resolve this problem we can make use of the identity  
\[
(\lambda^e\partial_e-\bar\lambda^{\dot e}\bar\partial_{\dot e}-\eta^{E}\partial_{E}+2)
\hat J=0.
\]
It holds by virtue of the definition \eqref{eq:Jhat} of $\hat J$
(total derivative)
and it represents the central charge condition $\gen{C}\hat J=0$.
This is also the reason why we started by acting on $\hat J$
representing the most general function with the property 
$\gen{C}\hat J=0$; our derivation only works for physical representations 
and the algebra closes only on when the central charge vanishes.
The derivatives of $\hat J_{1,2}$ w.r.t.\ $\alpha$ thus yield
\<\label{eq:dalphaJ}
\frac{d \hat J_1}{d\alpha}\eq
\cot\alpha(\lambda^e_1\partial_{1,e}+\bar\lambda^{\dot e}_1\bar\partial_{1,\dot e}+\eta^{E}_1\partial_{1,E})\hat J_1
=
\cot\alpha(2\lambda^e_1\partial_{1,e}+2)\hat J_1,
\nln 
\frac{d \hat J_2}{d\alpha}\eq
-\tan\alpha(\lambda^e_2\partial_{2,e}+\bar\lambda^{\dot e}_2\bar\partial_{2,\dot e}+\eta^{E}_2\partial_{2,E})\hat J_2
=
-\tan\alpha(2\lambda^e_2\partial_{2,e}+2)\hat J_2.
\>
Notice that the term without derivatives is sensitive to the number
of derivatives acting on $J$. For each fermionic derivative
the number $2$ is decreased by one unit.
Altogether we can write
\<
\ldots\eq
\pi^2 \varepsilon_{ac}\int \dd\alpha\,\frac{d}{d\alpha} 
\lrbrk{\sin\alpha\cos\alpha\bigacomm{\partial_{1,B}\hat J_1}{\partial_{2,D}\hat J_2}
-\sin^2\alpha\bigcomm{\hat J_1}{\partial_{2,B}\partial_{2,D}\hat J_2}}
\nln
\eq
-\pi^2 \varepsilon_{ac}
\bigcomm{\hat J(\Lambda)}{\partial_{B}\partial_{D}\hat J(0)}
=
\pi^2 \varepsilon_{ac}
\bigcomm{\partial_{B}\partial_{D}\hat J(0)}{\hat J(\Lambda)}.
\>
This has the form of a field-dependent gauge transformation
of the gauge covariant object $\hat J(\Lambda)$
because it maps $\hat J(\Lambda)\mapsto\comm{X}{\hat J(\Lambda)}$
where $X$ is the gauge variation parameter.

Finally we consider the anticommutator
of two correction terms $\acomm{({\gen S}_-)_{aB}}{({\gen S}_-)_{cD}}$.
We apply the sequence of two $\gen{S}_-$ to a source term $J(\Lambda)$
\<
(\gen{S}_-)_{cD}
(\gen{S}_-)_{aB}\hat J(\Lambda)\eq
4\pi^4\varepsilon_{ae}\varepsilon_{cf}\lambda^e\lambda^f
\int d\alpha\,d\beta\,\sin\alpha\, 
y \bigacomm{\comm{\hat J_y}{\partial_{x,D}\hat J_x}}{\partial_{z,B}\hat J_z}
\nlnum
-4\pi^4\varepsilon_{ae}\varepsilon_{cf}\lambda^e\lambda^f
\int d\alpha\,d\beta\,\sin\alpha\,
\frac{zy^2}{x^2+y^2} \bigcomm{\hat J_z}{\acomm{\partial_{y,B}\hat J_y}{\partial_{x,D}\hat J_x}}
\nl\nn
-4\pi^4\varepsilon_{ae}\varepsilon_{cf}\lambda^e\lambda^f
\int d\alpha\,d\beta\,\sin\alpha\,
\frac{zxy}{x^2+y^2}\bigcomm{\hat J_z}{\comm{\hat J_y}{\partial_{x,B}\partial_{x,D}\hat J_x}}.
\>
Note that for the latter two lines
we flipped the integration region $\alpha\mapsto\half\pi-\alpha$
in order to achieve a common parametrisation.
of $\Lambda_{x,y,z}$ where
$\lambda_x=x\lambda$, $\eta_x=x\eta$, etc., with 
\[
x= \sin\alpha\cos\beta,\qquad
y= \sin\alpha\sin\beta,\qquad
z= \cos\alpha.
\]
These are standard spherical coordinates and $d\alpha\,d\beta\,\sin\alpha$ is
the corresponding measure. The integral is over the positive octant,
$x,y,z>1$, such that we can freely exchange the coordinates $x,y,z$.
In the first line we exchange $y\leftrightarrow z$, multiply 
by $(x^2+y^2)/(x^2+y^2)$, and exchange $x\leftrightarrow y$ for the part
proportional to $x^2/(x^2+y^2)$. Upon use of a Jacobi identity 
on the second line the result reads
\<
(\gen{S}_-)_{cD}
(\gen{S}_-)_{aB}\hat J(\Lambda)\eq
4\pi^4\varepsilon_{ae}\varepsilon_{cf}\lambda^e\lambda^f
\int d\alpha\,d\beta\,\sin\alpha\, 
\frac{zy^2}{x^2+y^2} \bigacomm{\comm{\hat J_z}{\partial_{y,D}\hat J_y}}{\partial_{x,B}\hat J_x}
\qquad
\nlnum
-4\pi^4\varepsilon_{ae}\varepsilon_{cf}\lambda^e\lambda^f
\int d\alpha\,d\beta\,\sin\alpha\,
\frac{zy^2}{x^2+y^2} \bigacomm{\comm{\hat J_z}{\partial_{y,B}\hat J_y}}{\partial_{x,D}\hat J_x}
\nl\nn
-4\pi^4\varepsilon_{ae}\varepsilon_{cf}\lambda^e\lambda^f
\int d\alpha\,d\beta\,\sin\alpha\,
\frac{zxy}{x^2+y^2}\bigcomm{\hat J_z}{\comm{\hat J_y}{\partial_{x,B}\partial_{x,D}\hat J_x}}.
\>
This expression is manifestly symmetric in $a,c$, but manifestly antisymmetric 
in $B,D$. The anticommutator 
$\acomm{(\gen{S}_-)_{cD}}{(\gen{S}_-)_{aB}}$ thus vanishes.

In conclusion we find that 
$\acomm{{\gen S}_{aA}}{{\gen S}_{bB}}$
does not vanish for the interacting representation, 
but it closes onto a gauge transformation.
Our proof depended crucially on the 
assumption of vanishing central charge for all objects we act upon. 
Let us introduce the generator of a gauge transformation 
with gauge parameter $X$
\[\label{eq:gaugetra}
\gen{G}[X]=\pi^2\int d^{4|4}\Lambda \Tr\bigbrk{\comm{X}{J(\Lambda)} \check J(\Lambda)}.
\]
Our final result reads
\[\label{eq:S-S0}
\bigacomm{\gen{S}_{a A}}{\gen{S}_{bB}}
=
\eps_{ab}\gen{G}[\partial_A\partial_B J(0)].
\]

We now turn to the commutator of two generators $\bar {\gen S}$ acting on a source $\hat J(\Lambda)$.
We consider the anti-commutator of $\bar{\gen S}_0$ with $\bar {\gen S}_+$ yielding
\begin{align}
\bigacomm{(\bar{\gen S}_0)_{\dot a}^B}{(\bar {\gen S}_+)_{\dot b}^A}\hat J
=&-2\pi^2\int\dd^4\eta'\,\dd\alpha\,\eps_{\dot b\dot c}\eta_2^A
 \Big\{
 -\bar\lambda_1^{\dot c}\eta_1^B\bigcomm{\bar\partial_{1,\dot a}\hat J_1}{\hat J_2}\nln
&  -\bar\lambda_1^{\dot c} \eta_2^B\bigcomm{\hat J_1}{\bar\partial_{2,\dot a} \hat J_2}
 +\eta^B(\bar\partial_{\dot a}\bar\lambda_1^{\dot c})\comm{\hat J_1}{\hat J_2}
  +\eta^B\bar\lambda_1^{\dot c}\bar\partial_{\dot a}\comm{\hat J_1}{\hat J_2}
 \Big\}\,.
\end{align}
Using \eqref{eq:etadel1,eq:etadel2} this can be transformed to
\begin{multline}
\bigacomm{(\bar{\gen S}_0)_{\dot a}^B}{(\bar {\gen S}_+)_{\dot b}^A}\hat J
=-2\pi^2\int\dd^4\eta'\,\dd\alpha\,\varepsilon_{\dot b\dot c}\eta_2^A
 \Big\{
 -\bar\lambda_1^{\dot c}\eta'^B\cot\alpha\,\comm{\bar\partial_{\dot a}\hat J_1}{\hat J_2}\\
 +\bar\lambda_1^{\dot c}\eta'^B\tan\alpha\,\comm{\hat J_1}{\bar\partial_{\dot a}\hat J_2}
 +\delta_{\dot a}^{\dot c}\eta^B\sin\alpha\,\comm{\hat J_1}{\hat J_2}
 \Big\}\,.
\label{eq:acom1S0S+}
\end{multline}
The relevant term for the commutator
$\acomm{\gen{\bar{S}}_{\dot{a}}^B}{\gen{\bar{S}}_{\dot{b}}^A}$ is the
sum
\begin{equation}
 \bigacomm{\brk{\gen{\bar{S}}_0}_{\dot{a}}^B}{\brk{\gen{\bar{S}}_+}_{\dot{b}}^A}
+\bigacomm{\brk{\gen{\bar{S}}_+}_{\dot{a}}^B}{\brk{\gen{\bar{S}}_0}_{\dot{b}}^A}\,.
\label{eq:S0S+relevant}
\end{equation}
We split the commutator into its symmetric
and antisymmetric part
\begin{gather}
S^{BA}_{\dot a\dot b}=\bigacomm{\brk{\gen{\bar{S}}_0}_{\dot{a}}^{(B}}{\brk{\gen{\bar{S}}_+}_{\dot{b}}^{A)}}\,,
\qquad
A^{BA}=\varepsilon^{\dot{a}\dot{b}}\bigacomm{\brk{\gen{\bar{S}}_0}_{\dot{a}}^{[B}}{\brk{\gen{\bar{S}}_+}_{\dot{b}}^{A]}}\,,\nn\\
\bigacomm{\brk{\gen{\bar{S}}_0}_{\dot{a}}^B}{\brk{\gen{\bar{S}}_+}_{\dot{b}}^A}
+\bigacomm{\brk{\gen{\bar{S}}_+}_{\dot{a}}^B}{\brk{\gen{\bar{S}}_0}_{\dot{b}}^A}
=\varepsilon_{\dot{a}\dot{b}}A^{BA}+S^{BA}_{\dot a\dot b}\,,
\end{gather}
where 
\begin{equation}
X^{(AB)}=X^{AB}+X^{BA},\qquad X^{[AB]}=X^{AB}-X^{BA}.
\end{equation} 
Expanding $\lambda_1$ and $\eta_2$
according to \eqref{eq:Lambda12} and using antisymmetry under the shift
\begin{equation}
\alpha\mapsto\frac{\pi}{2}-\alpha\,,\qquad
\eta'\mapsto-\eta'\,,
\label{eq:varchangeS0S+}
\end{equation}
it is straightforward to show that the symmetric piece of \eqref{eq:S0S+relevant} vanishes.
The antisymmetric part reads
\begin{multline}
A^{BA}\hat J
=\pi^2\int\dd^4\eta'\,\dd\alpha\Big\{
 \sin\alpha\,(\eta_2^A\eta'^B-\eta_2^B\eta'^A)
\Big[
  \cot\alpha\,\bar\lambda^{\dot c}[\bar\partial_{\dot c}\hat J_1,\hat J_2]
 -\tan\alpha\,\bar\lambda^{\dot c}[\hat J_1,\bar \partial_{\dot c}\hat J_2]
 \Big]\\
 -2\sin\alpha\,(\eta_2^A\eta^B-\eta_2^B\eta^A)[\hat J_1,\hat J_2]
 \Big\}\,.
\end{multline}
Expanding $\eta_2$ and using (anti-)symmetry of some parts of the
integral under \eqref{eq:varchangeS0S+}, this can be written as
\begin{multline}
A^{BA}\hat J
=\pi^2\int \dd^4\eta'\,\dd\alpha
\Big\{-\eta'^A\eta'^B\bigbrk{\cot \alpha\,\bar\lambda^{\dot c}\comm{\bar\partial_{\dot c}\hat J_1}{J_2}
-\tan \alpha\,\bar\lambda^{\dot c}\comm{\hat J_1}{\bar \partial_{\dot c}\hat J_2}}\\
+(\eta'^A\eta^B-\eta'^B\eta^A)\comm{\hat J_1}{\hat J_2}\Big\}\,.
\label{eq:asymS0S+}
\end{multline}
We can now use an analogue of \eqref{eq:dalphaJ} for $\eta'\neq0$:
\begin{align}\label{eq:dalphaJeta}
\frac{d \hat J_1}{d\alpha}
&=
\cot\alpha(2\bar\lambda^{\dot e}_1\bar\partial_{1,\dot e}+2\eta_1^E\partial_{1,E}-2)\hat J_1
-\frac{1}{\sin\alpha\cos\alpha}\,\eta'^E\partial'_{E}\hat J_1,
\nln 
\frac{d \hat J_2}{d\alpha}
&=
-\tan\alpha(2\bar\lambda^{\dot e}_2\bar\partial_{2,\dot e}+2\eta_2^E\partial_{2,E}-2)\hat J_2
+\frac{1}{\sin\alpha\cos\alpha}\,\eta'^E\partial'_{E}\hat J_2.
\end{align}
Replacing $\bar \lambda^{\dot c}\bar\partial_{\dot c}$ in \eqref{eq:asymS0S+} 
by means of \eqref{eq:dalphaJeta} and making use of the identities
\begin{align}
&\int\dd^4\eta'\,\dd\alpha\,\eta'^A\eta'^B\eta^E
\bigbrk{\comm{\partial'_E\hat J_1}{\hat J_2}+\comm{\hat J_1}{\partial'_E\hat J_2}}
=-\int\dd^4\eta'\,\dd\alpha\,\eta'^{[A}\eta^{B]}\comm{\hat J_1}{\hat J_2},
\\
&\int \dd^4\eta'\,\dd\alpha\,\eta'^A\eta'^B\eta'^E
\bigbrk{\comm{\partial_E \hat J_1}{\hat J_2}-\comm{\hat J_1}{\partial_E\hat J_2}}
\nln&\qquad\qquad\qquad\qquad\qquad\qquad
=\int \dd^4\eta'\,\dd\alpha\,\eta'^A\eta'^B(\tan\alpha-\cot \alpha)\comm{\hat J_1}{\hat J_2},
\end{align}
we obtain
\begin{align}
A^{BA}\hat J
=-\pi^2\int\dd^4\eta'\,\dd\alpha\,\half\eta'^A\eta'^B\frac{d}{d\alpha}\comm{\hat J_1}{\hat J_2}
=\pi^2\int\dd^4\eta'\,\eta'^A\eta'^B\comm{J(0,\eta')}{J(\Lambda)}\,,
\end{align}
which amounts to a gauge transformation, cf.\ \eqref{eq:gaugetra}.

We refrain from explicitly calculating $\acomm{\gen{\bar{S}_+}}{\gen{\bar{S}_+}}$ 
since the result for $\acomm{\gen{\bar S}^A_{\dot a}}{\gen{\bar S}^B_{\dot b}}$ 
can alternatively be obtained by conjugation of $\acomm{{\gen S}_{aA}}{{\gen S}_{bB}}$:
\[
\acomm{\gen{\bar S}^A_{\dot a}}{\gen{\bar S}^B_{\dot b}}=
\eps_{\dot a\dot b}\gen{G}[\bar\partial^A\bar\partial^B \bar J(0)]
,
\]
where $\bar J$ is a complex conjugate source field
depending on conjugate odd variables $\bar\eta_A$.
The latter are related to the original odd variables $\eta^A$
through an odd Fourier transformation
(cf.\ \Secref{sec:MHVbar})
\[
\bar J(\bar\Lambda)=\int d^4\eta\, \exp(\eta^A\bar\eta_A)J(\Lambda).
\]
Converting back to the original source $J$ we obtain
\[
\bar\partial^A\bar\partial^B\bar J(0)=\int d^4\eta\, \eta^A\eta^B J(0,\eta)
=
\half\int \dd^4\eta\,\eta^A\eta^B \eta^C\eta^D\partial_C\partial_D J(0)
=
\half \varepsilon^{ABCD}\partial_C\partial_D J(0)
\]
and thus
\[\label{eq:S+S0}
\acomm{\gen{\bar{S}}_{\dot{a}}^A}{\gen{\bar{S}}_{\dot{b}}^B}
=
\half
\eps_{\dot a\dot b}\varepsilon^{ABCD}\gen{G}[\partial_C\partial_D J(0)].
\]

Finally we should mention that the inclusion of 
negative-energy particles discussed in \Secref{sec:NegEn}
leads to additional gauge transformation terms.

\subsection{Commutators involving \texorpdfstring{$\gen P$}{P} and \texorpdfstring{$\gen K$}{K}}

In order to evaluate the commutator of $\gen K$ and $\gen Q$ 
we can make use of $\gen K_{a\dot a}=\quarter\acomm{\gen S_{aB}}{\bar{\gen S}_{\dot a}^B}$ 
and the Jacobi identity to find
\[
\bigcomm{\gen K_{a\dot a}}{\gen Q^{bA}}=
-\quarter\bigcomm{\acomm{\bar{\gen S}_{\dot a}^B}{\gen Q^{bA}}}{\gen S_{aB}}
-\quarter\bigcomm{\acomm{\gen Q^{bA}}{\gen S_{aB}}}{\bar{\gen S}_{\dot a}^B}.
\]
The algebra of supercharges ensures that the first term 
vanishes and that the second term yields
\[
\bigcomm{\gen K_{a\dot a}}{\gen Q^{bA}}
=-\quarter\bigcomm{-\delta_a^b\gen{R}^A{}_B+\half\delta^b_a\delta_B^A\gen D}{\bar{\gen S}_{\dot a}^B}
=\delta_a^b \bar{\gen S}^A_{\dot a}.
\]
In other words this relation follows from consistency of the algebra
and there is nothing to be shown concerning the corrections to $\gen{K}$.
The commutators of $\gen K$ with $\bar{\gen Q}$ can be derived analogously 
\begin{equation}
\bigcomm{\gen K_{a\dot a}}{\bar{\gen Q}^{\dot b}_A}=\delta_{\dot a}^{\dot b}{\gen S}_{aA}.
\end{equation}
Finally, the commutator of $\gen{K}$ with $\gen{P}$ follows expressing the latter in terms of $\gen Q$ and $\bar{\gen Q}$ and employing the Jacobi identity 
\begin{equation}
 \bigcomm{\gen K_{a\dot a}}{\gen P^{b\dot b}}
=\quarter\bigcomm{\gen K_{a\dot a}}{\acomm{\gen Q^{bA}}{\bar{\gen Q}_A^{\dot b}}}
=\quarter\bigacomm{\gen Q^{bA}}{\comm{\gen K_{a\dot a}}{\bar{\gen Q}_A^{\dot b}}}
+\quarter\bigacomm{\bar{\gen Q}_A^{\dot b}}{\comm{\gen K_{a\dot a}}{\gen Q^{bA}}}.
\end{equation}
By means of the identities above this results in
\begin{equation}
\bigcomm{\gen K_{a\dot a}}{\gen P^{b\dot b}}
=\delta_{\dot a}^{\dot b}\gen L^b{}_a
+\delta_a^b\bar{\gen L}^{\dot b}{}_{\dot a}
+\delta_a^b\delta_{\dot a}^{\dot b}\gen D
\end{equation}
as expected.

For evaluating the commutator between $\gen{K}$ and $\gen{S}$, express
$\gen{K}$ in terms of $\gen{S}$ and $\gen{\bar{S}}$ and use the
Jacobi identity to find
\begin{equation}
\delta^C_A\comm{\gen{K}_{a\dot{a}}}{\gen{S}_{bB}}
=\bigcomm{\acomm{\gen{S}_{aA}}{\gen{\bar{S}}_{\dot{a}}^C}}{\gen{S}_{bB}}
=-\bigcomm{\acomm{\gen{S}_{bB}}{\gen{S}_{aA}}}{\gen{\bar{S}}_{\dot{a}}^C}
 -\bigcomm{\delta_B^C\gen{K}_{b\dot{a}}}{\gen{S}_{aA}}\,.
\label{eq:ksjacobi}
\end{equation}
By contracting once $C$ with $B$ and once $C$ with $A$ and taking a
linear combination, we obtain
\begin{equation}
\comm{\gen{K}_{a\dot{a}}}{\gen{S}_{bA}}
=\sfrac{1}{15}\bigcomm{\acomm{\gen{S}_{bB}}{\gen{S}_{aA}}-4\acomm{\gen{S}_{bA}}{\gen{S}_{aB}}}{\gen{\bar{S}}_{\dot{a}}^B}\,.
\end{equation}
Substituting the gauge transformation \eqref{eq:S-S0} 
\begin{equation}
\bigcomm{\gen{K}_{a\dot{a}}}{\gen{S}_{bA}}
=\sfrac{1}{3}
\eps_{ab}
\bigcomm{\gen{G}[\partial_A\partial_B J(0)]}{\gen{\bar{S}}_{\dot{a}}^B}
=
\eps_{ab}\gen{G}[\partial_A\bar\partial_{\dot a}J(0)].
\end{equation}
which amounts to a new gauge transformation $\gen{G}[\partial_A\bar\partial_{\dot a}J(0)]$.

For the commutator between $\gen{K}$ and $\gen{\bar{S}}$ one finds in
complete analogy with \eqref{eq:ksjacobi}
\begin{equation}
\delta^C_A\comm{\gen{K}_{a\dot{a}}}{\gen{\bar{S}}_{\dot{b}}^B}
=\bigcomm{\acomm{\gen{S}_{aA}}{\gen{\bar{S}}_{\dot{a}}^C}}{\gen{\bar{S}}_{\dot{b}}^B}
=-\bigcomm{\acomm{\gen{\bar{S}}_{\dot{a}}^C}{\gen{\bar{S}}_{\dot{b}}^B}}{\gen{S}_{aA}}
 -\bigcomm{\delta_A^B\gen{K}_{a\dot{b}}}{\gen{\bar{S}}_{\dot{a}}^C}\,.
\label{eq:ksbarjacobi}
\end{equation}
Again taking a linear combination of the two possible contractions of
this equation and using \eqref{eq:S+S0}, 
we obtain again a gauge transformation
\<
\bigcomm{\gen{K}_{a\dot{a}}}{\gen{\bar{S}}_{\dot{b}}^A}
\eq\sfrac{1}{15}\bigcomm{\acomm{\gen{\bar{S}}_{\dot{a}}^A}{\gen{\bar{S}}_{\dot{b}}^B}-4\acomm{\gen{\bar{S}}_{\dot{a}}^B}{\gen{\bar{S}}_{\dot{b}}^A}}{\gen{S}_{aB}}
=\sfrac{1}{6}
\varepsilon_{\dot{a}\dot{b}}\varepsilon^{ABCD}\comm{\gen{G}[\partial_C\partial_D J(0)]}{\gen{S}_{aB}}
\nln\eq
-\sfrac{1}{6}
\varepsilon_{\dot{a}\dot{b}}\varepsilon^{ABCD}
\gen{G}[\partial_a\partial_B\partial_C\partial_D J(0)]
\,.
\>

Finally, using the above results, we find that also
$\comm{\gen{K}_{a\dot{a}}}{\gen{K}_{b\dot{b}}}$ amounts to a gauge
transformation:
\<
\bigcomm{\gen{K}_{a\dot{a}}}{\gen{K}_{b\dot{b}}}
\eq\quarter\bigcomm{\gen{K}_{a\dot{a}}}{\acomm{\gen{S}_{bA}}{\gen{\bar{S}}_{\dot{b}}^A}}
=\quarter\bigacomm{\gen{\bar{S}}_{\dot{b}}^A}{\comm{\gen{K}_{a\dot{a}}}{\gen{S}_{bA}}}
 +\quarter\bigacomm{\gen{S}_{bA}}{\comm{\gen{\bar{S}}_{\dot{b}}^A}{\gen{K}_{a\dot{a}}}}
\nln\eq
\quarter 
\eps_{ab}\bigacomm{\gen{\bar{S}}_{\dot{b}}^A}{\gen{G}[\partial_A\bar\partial_{\dot a}J(0)]}
 +\sfrac{1}{24}
 \varepsilon_{\dot{a}\dot{b}}\varepsilon^{ABCD}
\bigacomm{\gen{S}_{bA}}{\gen{G}[\partial_a\partial_B\partial_C\partial_D J(0)]}
\nln\eq 
 \eps_{ab}
\gen{G}[\bar\partial_{\dot a} \bar\partial_{\dot b} J(0)]
+ \sfrac{1}{24}
 \varepsilon_{\dot{a}\dot{b}}\varepsilon^{ABCD}
\gen{G}[\partial_a\partial_b\partial_A\partial_B\partial_C\partial_D J(0)]
\>

To conclude we summarise the algebra relations closing onto gauge transformations
\eqref{eq:gaugetra}
\<
\acomm{\gen{S}_{a A}}{\gen{S}_{bB}}\eq
\eps_{ab}\gen{G}[\partial_A\partial_B J(0)],
\nln
\acomm{\gen{\bar S}^A_{\dot a}}{\gen{\bar S}^B_{\dot b}}\eq 
\half\eps_{\dot a\dot b}\varepsilon^{ABCD}\gen{G}[\partial_C\partial_D J(0)],
\nln
\comm{\gen{K}_{a\dot{a}}}{\gen{S}_{bA}}\eq
\eps_{ab}\gen{G}[\partial_A\bar\partial_{\dot a}J(0)],
\nln
\comm{\gen{K}_{a\dot{a}}}{\gen{\bar{S}}_{\dot{b}}^A}\eq -
\sfrac{1}{6}\varepsilon_{\dot{a}\dot{b}}\varepsilon^{ABCD}
\gen{G}[\partial_a\partial_B\partial_C\partial_D J(0)],
\nln
\comm{\gen{K}_{a\dot{a}}}{\gen{K}_{b\dot{b}}}\eq
\eps_{ab}\gen{G}[\bar\partial_{\dot a} \bar\partial_{\dot b} J(0)]
+
 \sfrac{1}{24}\varepsilon_{\dot{a}\dot{b}}\varepsilon^{ABCD}
\gen{G}[\partial_a\partial_b\partial_A\partial_B\partial_C\partial_D J(0)].
\>

The commutators of $\gen{P}$ with the supercharges follow 
analogously to the above commutators with $\gen{K}$.
Note that momentum conservation is not quite sufficient to show
the correct closure of these commutators.

\section{Exact Superconformal Invariance}
\label{sec:Trees}

\begin{figure}
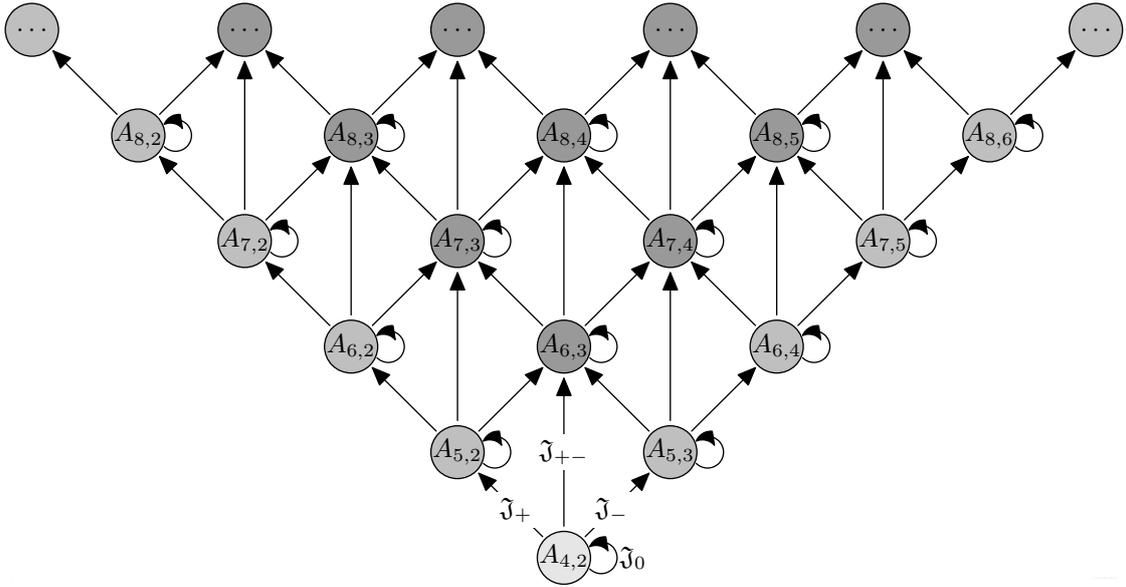
\centering
\includegraphicsbox[scale=1]{FigTreeChart.mps}
\caption{Illustration of the recursive action of the deformed generators.}
\label{fig:TreeChart}
\end{figure}

We would now like to extend the previous considerations, \Secref{sec:MHV} and \Secref{sec:MHVbar}, to the case of general tree amplitudes. 
We expect to find the obvious generalisation 
\[
\gen{J}_0 A_{n,k}+\gen{J}_+ A_{n-1,k}+\gen{J}_- A_{n-1,k-1}+\gen{J}_{+-} A_{n-2,k-1}=0.
\]
This gives rise to the pattern of relations shown in \Figref{fig:TreeChart} whereby 
a given amplitude is related to higher point amplitudes by the action of the deformed generators. 
As we have seen explicitly in the cases of MHV and $\overline{\rm MHV}$ amplitudes, 
the anomalous terms arise from 
collinear singularities seen by $\gen{J}_0$ which are then removed by $\gen{J}_+$
or  $\gen{J}_-$ as appropriate. In fact it is well known that the collinear behaviour 
is governed by the universal splitting functions and so we expect that the action of the
deformed generators is easily extended to the most general case. There are in
principle
contributions from other kinematic singularities which would need to 
be considered however none of these turn out to be
relevant for the action of the generators. 
We start our discussion with the concrete example of the six-point NMHV amplitude which
as we will see has, in addition to the collinear singularities, multi-particle poles
as well as apparent ``spurious'' (non-adjacent) singularities which
are non-physical and merely due to the methods for deriving the
expressions.

\subsection{Six-Point NMHV Amplitudes}
\label{sec:sixpoint}
For the case, ${A}_{6,3}={ A}\supup{NMHV}_6$,
that is to say, of six-point NMHV amplitudes,
we expect  that the action of ${\gen S}$
on the amplitude should be given by,
\[
\label{eq:sNMHV6}
\gen{S}_{0}{ A}\supup{NMHV}_6+\gen{S}_- {A}\supup{MHV}_5=0\,,
\]
where we note that the generator relates the six-point NMHV amplitude to 
the five-point MHV. 
We follow \cite{Drummond:2008vq} (see also \Appref{app:Conventions} for relevant definitions)
and write the six point NMHV amplitude as 
\[
{  A}^{\rm NMHV}_6={A}^{\rm MHV}_6(\half R_{146}+\mbox{cyclic})
\]
where there are several representations of $R_{146}$. One that is  particularly useful  
is
\[
R_{146}   ={ c}_{146}\deltad{4}(\Xi_{146})
\]
where
\<
c_{146}\eq\frac{\tprods{3}{4}\tprods{5}{6}}
{x_{14}^2\bra{1}x_{14}|4]\bra{3}x_{36}|6](\tprods{4}{5}\tprods{6}{1})^3[45][56]}\,,
\nln
\Xi^A_{146}\eq\tprods{6}{1}\tprods{4}{5}(\eta_4^A[56]+\eta_5^A[64]+\eta_6^A[45])\,,
\>
which is a specific example (after a little manipulation) of the general formula
\<
\label{eq:Rgen}
R_{pqr}\eq { c}_{pqr}\deltad{4}(\Xi_{pqr})\,,
\nln
c_{pqr}\eq\frac{\tprod{q-1}{q}\tprod{r-1}{r}}
{x^2_{qr}\bra{p}x_{pr}x_{rq-1}\ket{q-1}\bra{p}x_{pr}x_{rq}\ket{q}
\bra{p}x_{pq}x_{qr-1}\ket{r-1}\bra{p}x_{pq}x_{qr}\ket{r}}\,,
\nln
\Xi^A_{pqr}\eq-\bra{p}\Big[x_{pq}x_{qr}\sum_{i=p}^{r-1}\ket{i}\eta_i^A
+x_{pr}x_{rq}\sum_{i=p}^{q-1}\ket{i}\eta_i^A\Big].
\>

Now we want to consider the action of $\gen{S}$ on this amplitude 
and specifically the anomaly 
contribution coming from the action of $\partial$ on $1/\bar \lambda$ terms in 
the $R_{pqr}$ terms. As always one can use cyclicity to consider
a specific leg, for concreteness we
consider the $\bar \lambda_6$ terms. 
There are several different possible contributions to the anomaly terms:

\begin{enumerate}
\item
from multi-particle singularities which occur
when linear combinations of momenta such as ${(p_4+p_5+p_6)}$ become null. 
These singularities are of the form 
$\sum\tprods{j}{k}\ctprods{j}{k}$
and so do not contribute to the anomaly.%
\footnote{At tree level it is safe to assume 
a principal part prescription for propagators
and hence there are no further subtleties.}

\item
from singularities of the form $\bra{3}x_{46}|6]$ which occur when 
$p_4+p_5$ is any linear combination of $p_3$ and $p_6$. 
In fact these
singularities are spurious and cancel when we consider the full amplitude 
as can be explicitly seen in e.g.\ \cite{Mangano:1987xk,Mangano:1990by,Bern:1994cg,Bern:2007dw}. 
For a recent discussion of these
singularities in the twistor space approach see \cite{Hodges:2009hk}.

\item
collinear singularities due to $[56]$ type terms. 

\end{enumerate}

It is this last class that actually gives rise to the relevant physical 
singularities generating the anomaly terms
and that we will consider.
For completeness the full $R$ terms are
\<
\frac{1}{2}\left(R_{146}+R_{251}+R_{362}\right)\eq \frac{1}{2}\Bigg[ \frac{\tprods{3}{4}\tprods{5}{6}\tprods{6}{1}\tprods{4}{5}}
{x_{14}^2\bra{1}x_{14}|4]\bra{3}x_{36}|6][45][56]}
\deltad{4}\left(\eta_4[56]+\eta_5[64]+\eta_6[45]\right)
\nl
+ \frac{\tprods{4}{5}\tprods{6}{1}\tprods{1}{2}\tprods{5}{6}}
{x_{25}^2\bra{2}x_{25}|5]\bra{4}x_{42}|1][56][61]}
\deltad{4}\left(\eta_5[61]+\eta_6[15]+\eta_1[56]\right)
\nl
+ \frac{\tprods{5}{6}\tprods{1}{2}\tprods{2}{3}\tprods{6}{1}}
{x_{36}^2\bra{3}x_{36}|6]\bra{5}x_{53}|2][61][12]}
\deltad{4}\left(\eta_6[12]+\eta_1[26]+\eta_2[61]\right)\Bigg]
\nl
\>
and the anomaly term from the $[61]$  denominator factors 
in the second and third lines, and from the 
$[56]$ terms in the first and second lines give 
\<
(\gen{S}_{0})_{aB}{ A}\supup{NMHV}_6
\eq\frac{\pi}{2}\int\prod_{k=1}^6d^{4|4}\Lambda_k 
\Tr(\comm{J_6}{\partial_{1B}J_1}J_2J_3J_4J_5)
\nl
\frac{\deltad{4}(P_6)\deltad{8}(Q_6)}
 {\tprods{1}{2}\tprods{2}{3}\tprods{3}{4}\tprods{4}{5}\tprods{5}{6}}\deltad{2}(\tprods{6}{1})
\eps_{ab}\lambda_6^b
\nl
\left[\frac{\tprods{4}{5}\tprods{5}{6}\tprods{1}{2}}{x^2_{25}\bra{2}x_{35}|5]\bra{4}x_{42}|1][56]}\deltad{4}(\eta_6[15]+\eta_1[56])\right.
\nl
\left.+\frac{\tprods{5}{6}\tprods{1}{2}\tprods{2}{3}}{x^2_{36}\bra{3}x_{46}|6]\bra{5}x_{53}|2][12]}\deltad{4}(\eta_6[12]+\eta_1[26])\right]\,.
\>
Using manipulations identical to previous sections this can be rewritten as
\<
(\gen{S}_{0})_{aB}{A}\supup{NMHV}_6
\eq 2\pi^2 \int\prod_{k=2}^5d^{4|4}\Lambda_k d^{4|4} \Lambda'_1d\alpha d^4\eta' \ 
\frac{\deltad{4}(P'_5)\deltad{8}(Q'_5)}
 {\tprods{1'}{2}\tprods{2}{3}\tprods{3}{4}\tprods{4}{5}\tprods{5}{1'}}\deltad{4}(\eta')
\eps_{ab}\lambda_6^b
\nl
\kern+20pt\times\ 
\Tr(\comm{ \hat J_6 }{\partial_{1,B} J_1 }J_2J_3J_4J_5)
\,,
\>
where we have evaluated  the $\delta^2(\tprods{1}{6})$ and made use of the definitions
$\lambda_1=\lambda'_1\sin \alpha$, $\lambda_6=\lambda'_1\cos \alpha$,
$\eta_6=  \eta'_1 \sin \alpha+\eta'\cos \alpha$, $\eta_1= \eta'_1 \cos \alpha- \eta'\sin \alpha$.
This is consistent with $\gen{S}_-{\cal A}^{\rm MHV}_5$ using the expression \eqref{eq:s1+}
calculated from the action of
$\gen{S}$ on  ${\overline{\rm MHV}}$ amplitudes and thus we see that \eqref{eq:sNMHV6} does
indeed hold.
We now calculate the action of the undeformed generator $\bar{\gen{S}}$ on the six-point
NMHV amplitude. In this case we expect to find that 
\[
\label{eq:sbNMHV6}
\bar{\gen{S}}_{0}{ A}\supup{NMHV}_6+\bar{\gen{S}}_+{A}\supup{NMHV}_5=0\,.
\]
It is convenient to choose a slightly different writing of the six-point amplitude using the
formula \eqref{eq:Rgen}
\[
{ A}_n^{\rm NMHV}={ A}_n^{\rm MHV}\sum_{2\leq s,t\leq n-1}R_{nst}\,,
\]
where we sum over all $s$ and $t$ such that $s\neq t+1\ \rm{mod}\ n$. For the
specific case of six-points we take 
\[
{ A}_n\supup{NMHV}={ A}_n^{\rm MHV}(R_{624}+R_{625}+R_{635})
\]
and look for anomalous terms arising from the action of $\bar \partial$ on 
inverse powers of $\lambda$. We use cyclic symmetry to consider
only $\lambda_6$ and as in the previous case there are several 
possible sources for anomalous contributions, however, and
again as in the previous discussion only those singularities arising
from collinear singularities are relevant. Noting that  $R_{624}\sim \tprods{6}{1}$,
 $R_{625}\sim \tprods{6}{5} \tprods{6}{1}$ and $R_{635}\sim \tprods{6}{5}$ 
 we see that the only contribution from the singularity at $\lambda_6\propto\lambda_5$
 comes from the $R_{624}$ term and similarly the only contribution from the 
 $\lambda_6\propto\lambda_1$ singularity comes from the $R_{635}$ term. 
 Thus we find,
 \<
 (\bar{\gen S}_0)^A_{\dot a}{A}_6^{\rm NMHV}\eq-\pi \int \prod_{k=1}^6d^{4|4}\Lambda_k
 \Tr(J_1\dots J_6)\deltad{4}(P)\deltad{8}(Q) \eta_6^A
\nl
\times\left(\deltad{2}(\tprods{5}{6}) 
\frac{ \eps_{\dot a\dot b}{\bar \lambda}_5^{\dot b}R_{624}}{\tprods{6}{1}\tprods{1}{2}\dots\tprods{4}{5}}-\deltad{2}(\tprods{1}{6}) \frac{ \eps_{\dot a\dot b}{\bar \lambda}_1^{\dot b}R_{635}}{\tprods{1}{2}\dots\tprods{5}{6}}\right).
\nl
  \>
Evaluating the delta functions, using $\deltad{8}(Q')R_{1'24}=\deltad{8}(Q')R_{1'35}$, relabelling
the momenta and removing the phases we end up with
 \<
 (\bar{\gen S}_0)^A_{\dot a}{\cal A}_6^{\rm NMHV}\eq
-2\pi^2 \int \prod_{k=2}^5d^{4|4}\Lambda_kd^{4|4}\Lambda'_1d\alpha d^4\eta'
 \Tr(\comm{\hat{J}_6}{J_1}\dots J_5)
\nl \times \deltad{4}(P')\deltad{8}(Q') \eta_1^A\left(\frac{ \eps_{\dot a\dot b}{\bar \lambda}_6^{\dot b}R_{1'24}}{\tprods{1'}{2}\dots\tprods{5}{1'}}\right)
  \>
which is again consistent with the previous expressions for
$\bar{\gen{S}}_+$.

\subsection{General Tree Amplitudes and Splitting Functions}

It is useful to analyse the necessary behaviour of generic amplitudes so that our 
above results of \Secref{sec:sixpoint} generalise. As discussed, the important behaviour occurs
when two particles become collinear. For concreteness we consider the case
where
particle $n$ becomes collinear with particle $1$ with the scaling
\[
\lambda_n\rightarrow e^{i\varphi}\lambda_1' \sin \alpha,\qquad \lambda_1\rightarrow\lambda_1' \cos \alpha
\]
and the redefinitions
\[
 \eta_n=e^{-i \varphi}\eta'_1\sin \alpha+\eta' \cos \alpha,\qquad \eta_1=\eta'_1 \cos \alpha-e^{i \varphi}\eta' \sin \alpha.
\]
We postulate that a generic amplitude scales as
\<\label{eq:splitting}
{A}_{n,k}(\Lambda_1,\dots,\Lambda_n)\left.\right|_{1||n}\earel{\simeq}
\frac{e^{-i\varphi}\sec \alpha\csc\alpha }{\tprods{n}{1}}\,{ A}_{n-1,k}(\Lambda_1',\Lambda_2,\dots, \Lambda_{n-1})
\nl
+\frac{e^{i\varphi}\sec \alpha\csc\alpha }{[n1]}\,\deltad{4}(\eta'){ A}_{n-1,k-1}(\Lambda_1',\Lambda_2,\dots, \Lambda_{n-1})
\nl
+\mbox{finite terms},
\>
and with similar scaling in all other collinear limits.
Particular collinear limits of superspace
amplitudes were analysed in \cite{Drummond:2008cr} using the BCFW recursion 
relations described below.
Now assuming 
that the anomaly only receives contributions from the collinear singularities 
and that they scale as above it is straightforward to show that 
\<
({\bar {\gen S}}_0)_{\dot a}^A\mathcal{A}[J]\eq
-2\pi^2 \int \sum_n \prod_{k=2}^{n-1}d^{4|4}\Lambda_k d^{4|4}\Lambda'_1 \, d\alpha \, d^4\eta'\, 
\eps_{\dot a\dot b}{\bar \lambda}^{\dot b}_n \eta^A_1 { A}_{n-1, k}(\Lambda_1',\dots,\Lambda_{n-1})
\nl
\times\Tr(\comm{{\hat J}_n}{J_1}J_2\dots J_{n-1})
\>
and similarly 
\<
({ \gen S}_0)_{aA}\mathcal{A}[J]\eq
2\pi^2 \int \sum_n \prod_{k=2}^{n-1}d^{4|4}\Lambda_k d^{4|4}\Lambda'_1 \, d\alpha \, d^4\eta'\,
 \eps_{a  b}\lambda^{b}_n \deltad{4}(\eta') {A}_{n-1, k-1}(\Lambda_1',\dots,\Lambda_{n-1})
\nl
\times\Tr(\comm{\hat{J}_n}{\partial_{1,A}J_1}J_2\dots J_{n-1})
\>
which are both consistent with the expressions from the previous sections (as before we have removed the phases so that $\lambda_n=\lambda'_1\sin \alpha$, $\eta_n=\eta_1'\sin\alpha+\eta' \cos\alpha$ and by passing to the projection $\hat J_n$). 
\begin{figure}
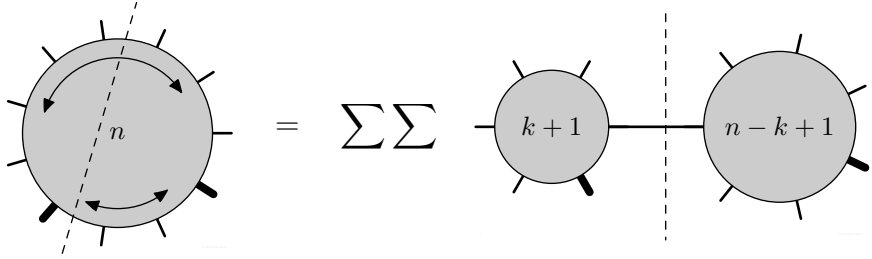
\centering
$\displaystyle
\includegraphicsbox{FigRecursionA.mps}
\quad=\quad \sum\sum\quad
\includegraphicsbox{FigRecursionB.mps}
$
\caption{Schematic of the on-shell recursion relation for a general tree-level amplitude. The shifted momenta are denoted by thickened legs and the sum is over the product of  subamplitudes which split
the shifted legs.}
\label{fig:BCFW}
\end{figure}

\begin{figure}
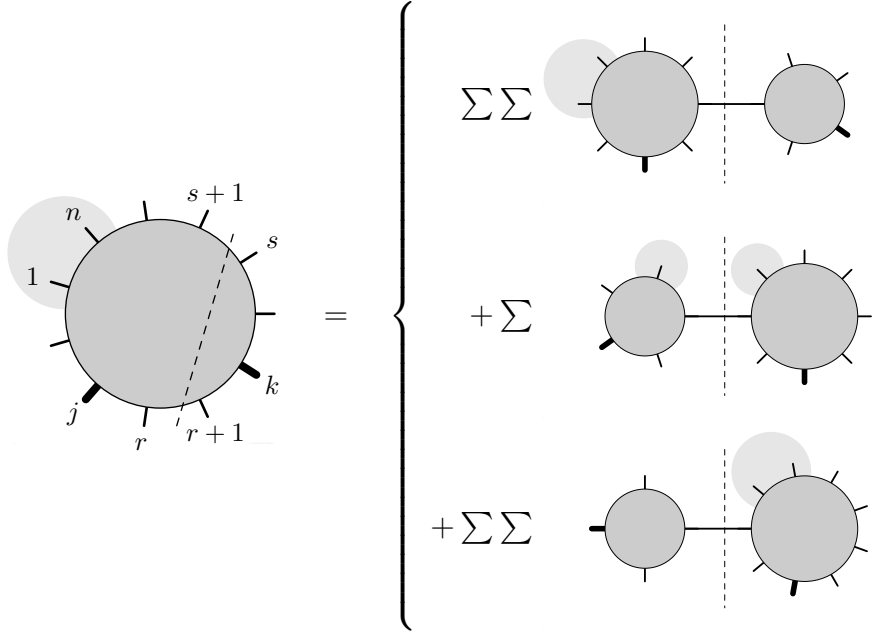
\centering
$\displaystyle
\includegraphicsbox{FigSplittingA.mps}\quad
=\quad
\left\{\begin{array}{r}
\sum\sum\includegraphicsbox[scale=0.7]{FigSplittingB.mps}
\\
+\sum\includegraphicsbox[scale=0.7]{FigSplittingC.mps}
\\
+\sum\sum\includegraphicsbox[scale=0.7]{FigSplittingD.mps}
\end{array}\right.
$

\caption{Illustration of the three positions of collinear legs in the recursion relations for the collinear legs 
chosen to be different than the shifted legs.}
\label{fig:BCFWColl}
\end{figure}

As previously mentioned, a convenient way to study arbitrary tree level amplitudes is to make use of the
BCFW recursion relations \cite{Britto:2004ap,Britto:2005fq} and for our purposes
the superspace versions \cite{Bianchi:2008pu,Brandhuber:2008pf,ArkaniHamed:2008gz,Elvang:2008na} are particularly useful. To verify the above
collinear structure \eqref{eq:splitting}, one does
not need to explicitly solve the recursion relations, as in \cite{Drummond:2008cr}, but can simply make
use of an inductive argument with the initial step being provided by the MHV and $\overline{\rm MHV}$ amplitudes 
considered previously. In the derivation of the BCFW relations one performs complex shifts of two of the
external legs, say $j$ and $k$,  and studies the resulting singularities. The resulting poles relate the amplitude to the
sum over products of subamplitudes with the momenta suitably shifted, shown schematically in \Figref{fig:BCFW}.
Following the usual procedure we shift 
$\hat{\tilde \lambda}_{j}={\tilde \lambda}_{j}+z_{rs}{\tilde \lambda}_k$,
and $\hat{ \lambda}_{k}={ \lambda}_{k}-z_{rs}{ \lambda}_{j}$; In the superspace
version we also shift the Gra{\ss}mann variables so that $\hat {\eta}_{j} =\eta_{j}+z_{rs}\eta_k$.
The resulting recursion relation can be written as 
\[
\label{eq:BCFW}
 A_{n}(\Lambda_1,\dots,\Lambda_n)=\int d^4P_{rs}\int d^4\eta_{rs} \sum_{r,s}{ A}_{L} \frac{1}{P_{rs}^2}{ A}_{R}.
\]
with%
\footnote{For this formula to be valid we must include in the sum three-point functions which are
non-vanishing for complex momenta. We must thus extend our earlier definition of $A_n$ to be 
$A_3=A_{3,1}+A_{3,2}$ for this special case.}
 \<
{ A}\indup{L}\eq A_{n-(s-r)+1}(\dots,\hat{\Lambda}_j,\dots,\Lambda_{r},\bar\Lambda_{rs},\Lambda_{s+1}),\nn\\
 {A}\indup{R}\eq A_{s-r+1}(\Lambda_{s},\Lambda_{rs},\Lambda_{r+1},\dots,\hat{\Lambda}_{k },\dots).
\>
The shift parameter $z_{rs}$ is determined by demanding the subamplitudes in each term to be on-shell. 
This is ensured to be the case if 
\[
(\hat{P}_{rs})^2=\lrbrk{\sum_{\ell=r+1}^{s}\lambda_\ell{\tilde\lambda}_\ell-z_{rs}\lambda_{j}{\tilde \lambda}_{k}}^2=0.
\]
We now want to consider the resulting behaviour as two legs become collinear and to show that
if all $n$-point amplitudes have the required behaviour, all the $(n+1)$-point amplitudes will too. 
This is simply a rewriting of the known universality of the splitting functions governing the 
collinear limit, \cite{Berends:1988zn,Mangano:1990by}, 
to the superspace notation via the BCFW recursion relations. 
In fact if the legs becoming collinear, again let us choose $n$ and $1$, 
are different than the shifted legs $j$ and $k$, it is easy to see that
the recursion relation \eqref{eq:BCFW} guarantees that this will be the case. 
There are three separate cases, as shown in \Figref{fig:BCFWColl}; 
when both collinear legs are on the left hand subamplitude $A\indup{L}$ 
which has the correct scaling by assumption, secondly when the collinear
legs are on different subamplitudes so there are no singularities and this term is subleading, finally 
when the two collinear legs are on the right hand subamplitude $A\indup{R}$
which again by assumption has the correct scaling.

\section{Conclusions and Outlook}
\label{sec:Concl}

In this paper we have considered superconformal invariance
of scattering amplitudes in $\superN=4$ SYM at tree level.
As the model is exactly superconformal, 
classically as well as quantum mechanically, 
observables ought to respect this symmetry. 
However, scattering amplitudes display 
collinear singularities which obscure the symmetries: 
At loop level they cause IR divergences 
which superficially break conformal symmetry. 
Further scrutiny reveals that collinear singularities
even break naive conformal symmetry at tree level.
This breakdown is easily overlooked because it only 
happens for singular configurations of the external momenta.
In order to understand the symmetries
of scattering amplitudes at loop level, it is crucial 
to first obtain complete understanding at tree level.

\smallskip

Here we have proposed to deform 
the \emph{free} superconformal generators $\gen{J}_0$ to 
\emph{classical interacting} generators, cf.\ \Figref{fig:NLin},
\[\label{eq:classgen}
\gen{J}=\gen{J}_0+\gen{J}_++\gen{J}_-+\gen{J}_{+-}.
\]
The correction terms cure the breaking of superconformal symmetry
at collinear singularities. 
They are what is known as non-linear realisations of symmetry;
as operators they act linearly, 
but they transform one field into several fields.
For scattering amplitudes it means that anomalous
terms in the action of the free generators are 
compensated by the interacting generators 
acting on amplitudes with \emph{fewer legs}.

\smallskip

We should note that the structure of singularities in
tree level scattering amplitudes is well understood. 
In general they correspond to internal propagators going on shell
meaning that the overall momentum of a subset of the external particles
becomes light-like. 
They can be classified into two-particle and multi-particle singularities:
Multi-particle singularities are codimension-one and do not lead 
to a conformal anomaly.
Conversely, two-particle singularities in Minkowski signature require 
the particles to be collinear. 
Collinearity is a codimension-two momentum configuration 
which leads to the conformal anomaly.
Collinear singularities can be expressed through
splitting functions times an amplitude with one leg less.
The conformal properties of splitting functions are understood.
It is also known how certain soft momentum limits of
the amplitudes are related to conformal symmetry.
Arguably our proposal constitutes 
a reformulation of what has been known 
about conformal symmetry for a long time.
In fact, the correction terms in \eqref{eq:classgen} 
can be understood as the action of the free conformal generators 
on the splitting functions \eqref{eq:splitting}. 
Nevertheless we believe that it is a useful formalisation
of classical conformal symmetry in view of extensions to the loop level.

Importantly we have shown that the deformations form a proper 
representation of $\alg{psu}(2,2|4)$ superconformal symmetry. 
Actually, the algebra does not close exactly but only modulo
field-dependent gauge transformations. 
This behaviour is not unexpected, it is rather very common in gauge field theories.
Here only the commutators of special superconformal generators 
$\gen{S},\bar{\gen{S}},\gen{K}$ yield gauge transformations. 
In a way this appears to be the dual of the very non-linear terms
in classical interacting gauge covariant supersymmetry transformations,
$\gen{Q},\bar{\gen{Q}},\gen{P}$. 
The latter act on the fields while our representation acts
on the dual sources noting that an algebra automorphism
maps between $\gen{S},\bar{\gen{S}},\gen{K}$ and $\gen{Q},\bar{\gen{Q}},\gen{P}$.

\medskip

An important insight is that conformal invariance
not only constrains the functional form of the amplitudes,
but it also constrains their singularities. 
In particular, invariance of the singularities 
requires cancellations between amplitudes 
with different numbers of legs,
cf.\ \Figref{fig:TreeChart}.
Hence, it does not make sense to consider an amplitude
with a fixed number of legs on its own, but only all amplitudes
at the same time, e.g.\ in the form of a generating functional
\eqref{eq:AmpGen}.
Therefore symmetry considerations 
can to some extent replace field theory computations 
which may become a very beneficial feature at higher loops.

\smallskip

Symmetries become even more powerful in the planar limit
where the superconformal algebra apparently extends to an
infinite-dimensional Yangian algebra.
Yangian symmetry leads to further constraints 
which prohibit certain superconformal invariants.
In fact, only a few invariants (up to anomalies) of the free Yangian 
are known \cite{Drummond:2008vq,Brandhuber:2008pf,Drummond:2009fd}. 
The tree level amplitude can be written as a linear
combination of these, but the coefficients are undetermined by symmetry.
Although we have not shown this explicitly, 
we are confident that full \emph{classical} Yangian symmetry 
(see \cite{Serban:2004jf,Agarwal:2004sz,Zwiebel:2006cb,Beisert:2007jv}
for interacting Yangians)
\[\label{eq:RepYangClass}
\genY{J}_\alpha=\half f_{\alpha}^{\beta\gamma}\sum_{1\leq k<\ell\leq n} \gen{J}_{k,\beta}\gen{J}_{\ell,\gamma},
\]
where $\gen{J}_{k,\beta}$ are the \emph{classical} generators in \eqref{eq:classgen},
leads to a \emph{unique} invariant which is precisely the 
tree scattering amplitude.
The point is that the naive invariants of the free Yangian 
have spurious singularities which are due to some decomposition 
of the amplitude into partial fractions. 
Physicality requirements can be used to argue for the
right linear combination. 
Our approach is different in that we merely rely on symmetry:
Spurious singularities are seen by the free generators,
but they are not cancelled by any interaction terms. 
Hence they should cancel among themselves leaving 
only the correct physical singularities.
In fact, unique determination of the tree level amplitude 
is an essential prerequisite for complete 
algebraic determination of loop amplitudes:
Tree-level invariants form the space of homogeneous solutions to the
covariance equations at loop level, i.e.\ they can be added freely to
loop amplitudes with arbitrary coefficients. If there is only a single
invariant, it must be the physical tree-level amplitude. Adding it to
the loop amplitude can be absorbed by changing the overall prefactor and
redefining the coupling constant, both of which cannot be determined by
algebraic means in any case. If there are multiple invariants, only
one of them can be identified with the tree-level amplitude and thus
the loop amplitude cannot be determined algebraically.

Note that we can easily argue for complete
Yangian invariance of the tree scattering amplitude.
According to \eqref{eq:RepYangClass}
the level-one momentum generator $\genY{P}$
(also known as the special dual conformal generator)
relies only on the superconformal generators 
$\gen{P},\gen{Q},\bar{\gen{Q}},\gen{L},\bar{\gen{L}},\gen{D}$.
All of these are free from holomorphic anomalies and receive no classical corrections,
thus $\genY{P}$ equals its free representation
for which invariance was shown in 
\cite{Drummond:2008vq,Brandhuber:2008pf,Drummond:2009fd}. 
All the other Yangian generators are obtained 
from commutators with superconformal generators.
Note that for completeness one should prove 
that \eqref{eq:RepYangClass} satisfies the Serre relations
of the Yangian algebra.
This would show that the closure of the algebra
generated by \eqref{eq:classgen,eq:RepYangClass}
is indeed a Yangian and not some other infinite-dimensional algebra.

Again, our interacting representation of the Yangian at tree level
does not add much to what is known already. 
It would demonstrate its full power only when quantum corrections are included: 
If there is a unique invariant at tree level, 
we expect the same to hold true at loop level. 
This would imply a complete determination of 
scattering amplitudes in planar $\superN=4$ SYM at all loops.
The price to be paid is the determination 
of corrections to the Yangian generators. 
This may or may not be simpler than determining the amplitude itself. 
Yet the formulation as a symmetry could ultimately enable 
certain non-perturbative statements, e.g.\ on the structure of singularities.

The possibility of a unique Yangian invariant scattering amplitude 
is also exciting for the spin chain point of view. 
When considered as a spin chain state, 
the scattering amplitude would be a representation of the unit operator
of the quantum mechanical spin chain model.
A Bethe ansatz based on this vacuum state could lead to 
a derivation of the exact spectrum of planar anomalous dimensions
alternative to the proposal in \cite{Bajnok:2008bm,Gromov:2009tv}
and follow-up works.

\bigskip

There are several issues deserving further investigation:

\smallskip

We did not consider conformal inversions in our work.
These can be used to define conformal boosts as 
shifts conjugated by conformal inversions. 
Free shifts do not receive classical corrections, 
consequently conformal inversions should carry those
corrections necessary for conformal boosts in this picture.
It is however not a priori guaranteed 
that conformal inversions are exact symmetries.
Are scattering amplitudes invariant under the 
superconformal group including inversions or 
merely under the component connected to the identity?

\smallskip

It would be desirable to prove that classical Yangian symmetry 
determined the tree scattering amplitude uniquely. 
Can one show that there is only a single invariant?

\smallskip

The proposed corrections to superconformal symmetry 
are based on the holomorphic anomaly 
which requires a spacetime with $(3,1)$ Minkowski signature.
Many works on tree level scattering amplitudes 
make use of a twistor transform which is most
conveniently defined in $(2,2)$ signature. 
It would be interesting to find out whether our results
can also be formulated for this split signature. 
Clearly, the holomorphic anomaly would have to be 
replaced by something else. 
One could contemplate postulating the equivalent of \eqref{eq:spinoranomaly}. 
Alternatively one could try to find different anomalous terms in the
action of the free generators. In the spinor helicity framework it 
is not immediately clear how to define such terms 
but in the twistor space representation 
the various signum factors \cite{Mason:2009sa,ArkaniHamed:2009si} do give
rise to singular contributions when two spinors become collinear. 
Cancellations then might involve also three-leg and two-leg amplitudes
in this signature.
Moreover, we expect that N$^{-1}$MHV amplitudes
would play a role; like the three-leg amplitudes
these have a restricted support in momentum space.

\medskip

The interacting representation of superconformal symmetry
does not rely on the planar limit or on integrability 
and therefore one may wonder if similar formulations can be obtained for
field theories with less supersymmetry. 
In particular, all tree scattering amplitudes
in pure $\superN<4$ supersymmetric gauge theories 
(including pure Yang--Mills at $\superN=0$)
equal the restriction of the $\superN=4$ counterparts.
Also the truncation of the classical $\alg{psu}(2,2|4)$ representation 
to $\alg{su}(2,2|\superN)$ is consistent. 
It is a proper representation 
that annihilates all truncated amplitudes. 
This appears to work independently of the conformal anomaly 
at one loop due to a non-trivial beta-function. 
It is however not immediately clear whether one can add massless matter 
to $\superN<4$ field theories
and still obtain a proper representation of conformal symmetry
which annihilates all tree amplitudes.

\smallskip

Finally, we would like to mention the possibility of
establishing a similar framework for $\superN=8$ supergravity.
In this model the $\grp{E}_{7(7)}$ global symmetry has features
reminiscent of the special conformal symmetries
including relations between amplitudes with different numbers of 
legs, the behaviour in collinear and soft limits
(see e.g.\ \cite{ArkaniHamed:2008gz})
as well as the structure of generators and their algebra,
(see e.g.\ \cite{Brink:2008qc,Kallosh:2008ic}).

\paragraph{Acknowledgements.}

We are grateful to
Nima Arkani-Hamed, 
Lars Brink, 
James Drummond,
Johannes Henn, 
Jared Kaplan,
Lionel Mason, 
Jan Plefka,
Radu Roiban,
David Skinner,
Emery Sokatchev 
and
Matthias Staudacher 
for interesting discussions and for sharing useful insights.
N.B.~thanks the IPPP Durham 
for hospitality during the workshop ``Amplitudes 09''
which provided the inspiration for this work. 
T.B., N.B.\ and F.L.\
would like to thank the Galileo Galilei Institute for Theoretical Physics
for hospitality during the workshop ``New Perspectives in String Theory''
where part of this work was carried out.

\appendix

\section{Conventions}
\label{app:Conventions}

  \begin{itemize}
  \item We will mostly consider the $(3,1)$ signature $(\mathord{-}\mathord{+}\mathord{+}\mathord{+})$. The positive and negative chirality spinors 
  are denoted by $\lambda^a$, $a=1,2$ and ${\tilde\lambda}^{\dot a}$, $\dot a=1,2$. 
  \item We have for the antisymmetric two tensor: $\eps_{12}=-\eps_{21}=1$ and $\eps^{21}=-\eps^{12}=1$
  so that $\eps^{ab}\eps_{bc}=\delta_c^a$. 
The antisymmetric four tensor $\eps^{ABCD}$ is defined such that $\eps^{1234}=\eps_{1234}=+1$.
  \item We define the positive chirality spinor brackets 
  $\tprod{\lambda_1}{\lambda_2}=\eps_{ab}\lambda_1^a\lambda_2^b$  and the negative
  chirality brackets  $\comm{{\tilde \lambda}_1}{{\tilde \lambda}_2}=\eps_{\dot a\dot b}{\tilde \lambda}_1^{\dot a}{\tilde \lambda}_2^{\dot b}$. The same conventions apply for the abbreviations $\tprods{i}{j}$ and $[ij]$. 
  \item A four-vector $p_\mu$ can be thought of as a bi-spinor 
  $p^{a\dot a}=(\sigma^\mu)^{a\dot a}p_\mu$ which for light-like vectors can be written as
  $p^{a\dot a}=\lambda^a{\tilde \lambda}^{\dot a}$ for some spinors
  $\lambda$, $\tilde\lambda$. In $(3,1) $
  signature demanding that $p_\mu$ be real implies that ${\tilde \lambda}=\pm \bar \lambda$. 
  When $p_\mu$ is a particle four-momentum, the sign corresponds to positive and 
  negative energy. 
  \item It is useful to introduce the dual variables $(x_i)^{ a\dot a}$, $i=1,\dots,n $ defined by $x_i-x_{i+1}=p_i$
  satisfying the condition $x_{n+1}=x_1$. We make use of the shorthand 
  $x_{rs}=x_r-x_s=\sum_{i=r}^{s-1}p_i$. As well as
  \<
  \bra{p}x_{mn}|q]&=&\lambda_p^a(x_{mn})_{a\dot a}{\tilde \lambda}_q^{\dot a}\nn\\
   \bra{p}x_{mn}x_{kl}\ket{q}&=&\lambda_p^a(x_{mn})_{a\dot a}(x_{kl})^{\dot a b}\eps_{bc}\lambda_q^c
   \>
  \item For treating complex variables the convention for the measure is 
 $d^2 z=dxdy$ where $ z=x+i y$. We define derivatives $\partial$ and ${\bar \partial}$ so that
$\partial z=1$ and ${\bar \partial}z=0$ etc. We also define 
 \[
 \int d^2 z \ \deltad{2}(z)=1
 \]
 so that $\deltad{2}(z)=\delta(x)\delta(y)$. This implies for the holomorphic anomaly that%
\footnote{For distributions and this particular relation see e.g.\ \cite{Gelfand:1964aa}.}
 \[
\frac{\partial}{\partial \bar z}\,\frac{1}{z}=\pi \deltad{2}(z).
 \]
In other words, $1/z$ is the Green's function for the differential operator
$\partial/\partial\bar z$.
This can be easily seen, and the overall coefficient fixed, by making use of Green's theorem
\[
\int_{\mathcal{R}} d^2z\ \frac{\partial}{\partial \bar z}\, \frac{1}{z}=-\frac{i}{2}\oint_{\partial \mathcal{R}} dz\ \frac{1}{z}\,.
\]
\item
We assume that we are in $(3,1)$ signature and we treat the $\lambda^a$'s as 
  complex variables so that $d^4\lambda=d^2\lambda^1\,d^2\lambda^2$. 
  In particular it is defined so that 
\begin{equation}
\deltad{2}(\tprod{\lambda}{\mu})=\int d^2z\,\deltad{4}\brk{\lambda-z\mu}
\label{eq:deltatrans}
\end{equation}
and
  \[\label{eq:deltabracketeval}
  \int d^4\lambda \, \deltad{2}(\tprod{\lambda}{\mu})f(\lambda, \bar{\lambda})
=  \int d^4 \lambda\, d^2z\, \deltad{4}(\lambda - z \mu)f(\lambda,\bar{\lambda})
=
  \int d^2z\, f(z \mu,\bar z\bar{\mu}).
   \]

 \item Gra{\ss}mann integration is defined as $\int d\eta=0$ and $\int d\eta\ \eta=1$.
The odd delta function is consequently defined as $\delta(\eta)=\eta$.
Integral over all four $\eta^A$'s is defined as
$d^4\eta=d\eta^1\,d\eta^2\,d\eta^3\,d\eta^4$
and the odd delta function such that $\int d^4\eta\,\deltad{4}(\eta)=1$.

\item The superspace integration measure, $d^{4|4}\Lambda$, is defined to be
$d^{4|4}\Lambda=d^4\lambda\, d^4\eta$.
  \end{itemize}


\bibliographystyle{nb}
\bibliography{confscat}

\end{document}